\documentclass[11pt]{article}
\usepackage{latexsym}
\usepackage{graphicx}
\usepackage{multirow}
\usepackage{amsmath}
\usepackage[square,sort&compress]{natbib}
\usepackage{setspace}
\usepackage{subfigure}
\begin{document}
\begin{center}\Large{Electrostatic-gravitational oscillator}
\end{center}
\begin{center}
\renewcommand\thefootnote{\fnsymbol{footnote}}
Constantinos G. Vayenas \footnote{E-mail: cgvayenas@upatras.gr} \& Stamatios Souentie
\end{center}
\begin{center}
\textit{LCEP, Caratheodory 1, St., University of Patras, Patras GR 26500, Greece}
\end{center}
\begin{abstract}
{We examine the one-dimensional motion of two similarly charged particles under the influence of only two forces, i.e. their Coulombic repulsion and their gravitational attraction, using the relativistic equation of motion. We find that when the rest mass of the two particles is sufficiently small $(\sim 0.4{\rm \; }eV/c^{2})$ and the initial Coulombic potential energy is sufficiently high ($\sim m_{p}c^{2}$, where $m_{p}$ is the proton mass), then the strong gravitational attraction resulting from the relativistic particle velocities suffices to counterbalance the Coulombic repulsion and to cause stable periodic motion of the two particles. The creation of this confined oscillatory state, with a rest mass equal to that of a proton, is shown to be consistent with quantum mechanics by examining the particle de Broglie wavelength and the Klein-Gordon and Schr\"odinger equations. It is shown that the gravitational constant can be expressed in terms of the proton mass and charge, the vacuum dielectric constant, the Planck constant and the speed of light. It is also shown that gravity can cause confinement of light $(\sim 0.1{\rm \; }eV/c^{2})$ neutral particles (neutrinos), or pairs of a neutral and a charged light particle, in circular orbits of size 0.9 $fm$ and period $10^{-24}$ $s$ forming bound neutral or charged hadron states.} 
\end{abstract}
\newpage
\section{Introduction}
The possibility that  gravity may have a significant role at short, femtometer or subfemtometer distances has attracted significant interest for years \cite{Hoyle01,Long03,Antoniadis98,Randall99,Pease01,Abbott01,Hehl80} and the potential role of special relativity \cite{Schwarz04} as well as the feasibility of developing a unified gauge theory of gravitational and strong forces \cite{Hehl80} have been discussed. 

Gravitational forces between small particles at rest are entirely negligible in comparison with Coulombic forces. Thus the gravitational attraction between two protons at rest is 36 orders of magnitude smaller than their Coulombic repulsion, i.e. $Gm_{p}^{2} =8.03\cdot 10^{-37} {\rm \; (e}^{{\rm 2}} /\varepsilon )$ where we denote $\varepsilon =4\pi \varepsilon_{o} $.

Due to this enormous 36 orders of magnitude gap, little attention has been focused on gravitational forces between fast moving particles with velocities very close to c. For the laboratory observer in frame S (Figure 1) the one-dimensional relativistic equation of motion of a particle with rest mass $m_{o} $ is \cite{{French68},{Freund08}}:

\begin{equation}
\label{eq1}
F=\frac{dp}{dt}=\frac{d(\gamma m_{o}v)}{dt}=\gamma ^{3}m_{o}\frac{dv}{dt}=m_{\ell}\frac{dv}{dt}
\end{equation}
\begin{figure}[h]
\centerline{\includegraphics[width=9.80cm,height=11.00cm]{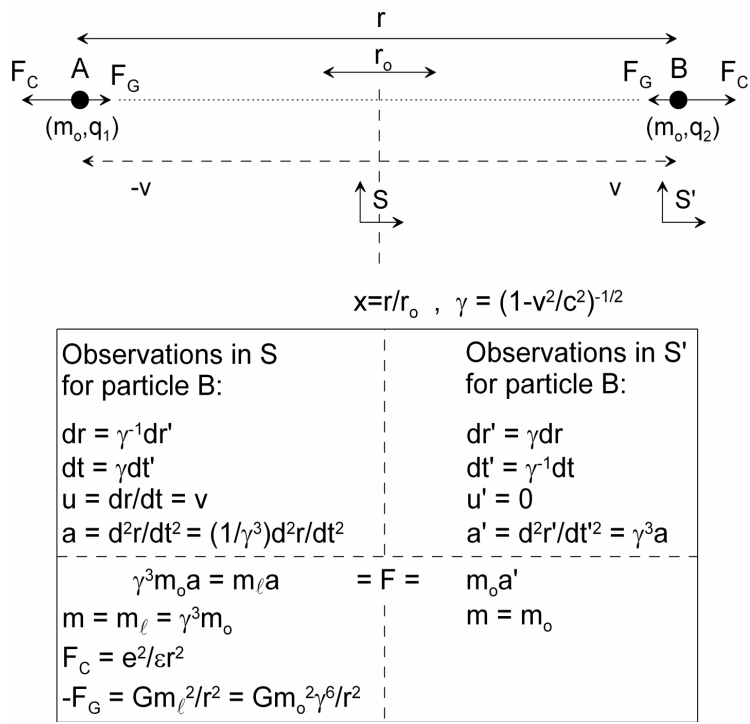}}
\caption{Schematic of the two charged particles with rest mass $m_{o} $, of the symmetry axis and of the two reference frames $S$ and $S'$. Inset summarizes the key \cite{French68,Freund08} relativistic relationships between the observations in frames $S$ and $S'$; $\gamma ^{3} (r)m_{o} $ is the longitudinal mass \cite{Freund08} at displacement $r$; The same force value, F, is observed in both frames, but the two observed accelerations, $a'$ and $a$, have a ratio of $\gamma ^{3} $ \cite{French68,Freund08}.}
\label{fig1}
\end{figure}

\noindent {where $m_{\ell}=\gamma ^{3} m_{o} $ is the longitudinal mass of the particle, $\gamma(=(1-v^{2}/c^{2})^{-1/2})$ is the Lorentz factor and $v=$v is the velocity of the particle relative to the laboratory observer \cite{French68,Freund08}. As (\ref{eq1}) shows, $m_{\ell } $ is the ratio of force divided by acceleration, thus it is the inertial mass of the particle \cite{Schwarz04,French68,Freund08}, i.e. the quantity defined and measured as the mass of all bodies subject to gravity \cite{French68,Freund08}. It is on the basis of this quantity that Newton's $r^{-2} $ law has been formulated and Newton's constant has been measured. In view of the well proven equivalence principle \cite{Schwarz04} it is thus equal to the gravitational mass in Newton's $r^{-2}$ gravitational law \cite{Schwarz04}. It is also the only mass value for the moving particle B which the laboratory observer $S$ can judge on the basis of his force and acceleration observations, the rest mass $m_{o} $ of the moving particle cannot be measured in frame $S$, only $m_{\ell } $ is measurable in the laboratory frame. Thus it is this quantity, rather than the rest mass $m_{o} $, which is available to the laboratory observer to use in Newton's gravitational law applied to the one-dimensional motion of the moving particle, i.e.} 

\begin{equation}
\label{eq2} 
F_{G}=\frac{dp}{dt} =m_{\ell } \frac{dv}{dt} =\frac{Gm_{\ell }^{2}}{r^2}=\frac{Gm_{o}^{2} \gamma ^{6}}{r^2}
\end{equation} 

For particles with non-relativistic velocities this  subtle difference is negligible. Also at a first glance the $\gamma ^{6} $ term appears insufficient to make $F_{G} $ significant relative to Coulombic forces unless the velocity v is high enough to bring the Lorentz factor $\gamma $ close to $10^{6} $. However such relativistic velocities can in principle be easily reached by small particles under the influence of their Coulombic attraction or repulsion.

Thus in this work we examine the seemingly very simple problem of the one-dimensional motion of two similarly charged small particles under the influence of only two forces, i.e. their Coulombic repulsion and gravitational attraction using the relativistic equation of motion and the Coulomb and Newton laws in order to examine under what conditions the two particles will not simply escape from each other, but may collide with each other or form a stable oscillatory state. At the end we show the consistency of the results with quantum mechanics by examining the de Broglie wavelengths of the two particles and the Klein-Gordon equation of quantum mechanics which allows for the use of the Schr\"odinger equation under relativistic conditions. 

We focus on examining the conditions for which the rest mass of the oscillatory two-particle system corresponds to that of a hadron such as a proton. There has been some preliminary work in this area treating hadrons as standing waves \cite{Vayenas07} or strings \cite{Vayenas2007} and leading to the same analytical expression for G derived here, but the exact origin of the creation of the standing wave bound state was unclear. The formation of hadrons via condensation of smaller particles, i.e. of the gluon-quark plasma, is commonly analyzed in the QCD theory \cite{Braun07,Gross73, Politzer73, Cabibbo75, Aoki06, Fodor04}. This condensation occurs at the transition temperature of QCD which is given by $T_{C}=151(6)$ MeV in the kT scale \cite{Aoki06}, i.e. it corresponds to a particle energy of approximately 150 MeV \cite{Aoki06}. 

\section{Reference frames}
As is common practice \cite{French68,Freund08}, we examine the motion of each particle (e.g. particle B) using two reference frames (Figure 1). The laboratory frame S is at zero velocity with respect to the center of mass of the two particles, while the instantaneous rest frame $S'$ has a velocity with respect from S equal to that of particle B \cite{French68,Freund08}. 

The inset of Figure 1 summarizes the basic elements of special relativity \cite{French68,Freund08} regarding the values of force, velocity and acceleration of particle B observed in the two reference frames. The only one of these basic elements needed for the present analysis is that, although the force value, F, observed in both reference frames is the same \cite{French68,Freund08}, the acceleration, a, measured by the laboratory observer $S$ is $\gamma ^{3} $ times smaller than the acceleration, $a'$, measured in frame  $S'$ and thus the mass value $m_{\ell } $ judged by the laboratory observer is $\gamma ^{3} $ times larger than the mass $m_{o} $ judged in frame $S'$, which is the rest mass $m_{o} $ of the particle since the particle is at rest $(u'=0)$ in frame $S'$ \cite{French68,Freund08}. Thus the longitudinal mass $m_{\ell } =\gamma ^{3} m_{o} $ is the only mass value observed by the laboratory observer and thus, in view of the equivalence principle \cite{Schwarz04}, the only one available to use in Newton's gravitational law for the one-dimensional particle motion, as already discussed. The rest mass $m_{o} $ of the accelerating particle is not measurable by the laboratory observer. This simple but key point is worth emphasizing because what follows after equation (\ref{eq2}) in treating the stated one-dimensional particle motion problem is then basically simple energy and momentum conservation and the corresponding simple algebra. 

\section{Initial conditions}
We denote $r_{o}$ the initial particle distance and we define $x=r/r_{o}$. Since we are interested in the possible formation of a hadron with charge e, i.e. a proton, we consider due to charge conservation, that the charges of the two particles of rest mass $m_o$ each are $q_{1} $ and $q_{2} $ with $q_{1} +q_{2} =e$. Denoting $q_{1}/e=Q_{1} $ and $q_{2} /e=Q_{2} $ and $Q_{1} Q_{2}=Q$ $(\le 1/4)$ it follows that the initial Coulombic potential, $V_{C,o} $, is given by:  

\begin{equation}
\label{eq3} 
V_{C,o}=\frac{Qe^{2}}{\varepsilon r_{o}}
\end{equation}

 We denote: 

\begin{equation}
\label{eq4} 
\rho=\frac{V_{C,o} }{m_{p}c^{2}}
\end{equation} 

where $m_{p} $ is the proton mass, thus: 

\begin{equation}
\label{eq5} 
r_{o}=\frac{Qe^{2}}{\rho \varepsilon m_{p}c^{2}}=\left(\frac{Q}{2\rho}\right)\lambda_{q}=\left(\frac{Q}{\rho}\right)\lambda_{c}(\alpha/2\pi) 
\end{equation} 

\noindent {where $\lambda_{c}(=h/m_{p} c=1.32{\rm \; fm)}$ is the proton Compton length, $\alpha(=e^2/\epsilon c\hbar=1/137.035)$ is the fine structure constant (so that $\alpha/2\pi=e^2/\epsilon ch)$, and the length $\lambda_{q}(=2e^{2}/\varepsilon m_{p} c^{2} =2\lambda _{c} (\alpha /2\pi )=3.06\cdot 10^{-3} {\rm \; fm)}$ is of the size range of quarks (thus we use the subscript q) and is twice what is commonly termed ``classical radius'' of the proton.} 

Since $x=r/r_{o} $ it follows from (\ref{eq5}) that:

\begin{equation}
\label{eq6} 
x=\left(\frac{2\rho }{Q} \right)\left(\frac{r}{\lambda_{q}}\right) 
\end{equation} 

Thus the parameter $\rho (=V_{C,o}/m_{p}c^{2})$ defines via (\ref{eq5}) the initial $(t=0)$ particle distance $r_{o} $ and thus the initial dimensionless distance $x_{o}(=1)$. 

In choosing the initial velocity, $v_{o}$, or equivalently the initial kinetic energy $K_{o}(=(\gamma_{o}-1)m_{o}c^{2})$ of each particle, we have examined two general cases:
(a) $2K_{o}=(1-\rho)m_{p} c^{2} $, so that $2K_{o} +V_{C,o} =m_{p} c^{2} $, which we denote case A. 
(b) $K_{o} =0$, thus $\gamma _{o}=1$, which we denote case B.

The two cases coincide for $\rho =1$.

\section{Energy conservation}
If due to the action of gravity the two particles form a bound state then the total energy $2E(x)=2\gamma(x)m_{o}c^{2}$ of the two initial particles becomes the rest energy, ${R(x)}$, of the confined state. In this case the two particles can be viewed as partons of the confined hadron state in Feynman's parton model or, as shown later, as quarks and gluons in K. Johnson's bag model or in the standard model. If the rest mass $R(x)$ is at some point $x_{p}$ equal to the rest energy of a proton, i.e. $R(x_p)=m_{p} c^{2} $, then denoting $\gamma (x_{p} )=\gamma _{p} $ it is:

\begin{equation}
\label{eq7} 
m_{p}=2\gamma_{p}m_{o}{\qquad};{\qquad}\gamma_p=m_{p}/2m_{o}
\end{equation} 

It is worth noting that since in general the total particle energy $E(x)$ is related to its momentum $p(x)$ via:

\begin{equation} 
\label{eq8} 
E^{2}(x)=m_{o}^{2}c^{4}+p^{2}(x)c^{2}
\end{equation}
 
and since $R(x)=2E(x)$ it follows:

\begin{equation} 
\label{eq9} 
R^{2}(x)=4m_{o}^{2}c^{4}+4p^{2}(x)c^{2}
\end{equation}

We start the derivation by using energy conservation:

\begin{eqnarray}
\label{eq10}
R(x)=2\gamma(x)m_{o}c^{2}=2m_{o}c^{2}+2\left(\gamma(x)-1\right)m_{o}c^{2}=2m_{o} c^{2} +2K(x)\nonumber\\ 
=2m_{o} c^{2} +2K_{o}+V_{C,o}+V_{G,o}-V_{C}(x)-V_{G}(x)
\end{eqnarray} 
where $K_{o} $ is the initial kinetic energy of each particle, $V_{C,o} $ and $V_{G,o} $ are the Coulombic and gravitational potential energies at $x=1$ (i.e. at $r=r_{o} $), $V_{C} (x)(=V_{C,o} /x)$ and $V_{G} (x)$ are the Coulombic and gravitational potential energies at x  and $K(x)$is the kinetic energy of each particle at x. 

Choosing to analyze first the case of initial conditions A, i.e. $2K_{o}=(1-\rho )m_{p} c^{2} $ so that $V_{C,o} +2K_{o} =\rho m_{p} c^{2} +(1-\rho )m_{p} c^{2} =m_{p} c^{2} $  and dividing (\ref{eq10}) by $2\gamma _{p} m_{o} c^{2} =m_{p} c^{2} $ one obtains:

\begin{equation}
\label{eq11} 
\gamma (x)/\gamma _{p}=\mu (x)=1-\rho \left(\frac{1}{x} -y(x)\right) 
\end{equation}
 
where we have accounted for $V_{C,o} +K_{o} =m_{p} c^{2} $, for $V_{C} (x)=V_{C,o} /x$ and for $m_{o} \ll m_{p} $, i.e. for $\gamma _{p} \gg 1$ which is the case as shown later, and where we have defined $y(x)=-(V_{G}(x)-V_{G,o})/V_{C,o} $ and $\mu (x)=\gamma (x)/\gamma _{p} =R(r)/m_{p} c^{2} $.

 Note that for zero relative velocity of the two particles it is $y(x)=\xi /x$, where $\xi=\varepsilon Gm_{p}^{2}/e^{2}\approx 10^{-36}$, but for relativistic velocities the function $y(x)$ becomes significant and has to be determined (section 4.1). 

 It follows from (\ref{eq11}) that when $\mu (x)=1$ or, equivalently, $R(x)=m_{p} c^{2} $ then:

\begin{equation}
\label{eq12} 
y(x_{p} )=1/x_{p} {\qquad};{\qquad}-(V_{G}(x)-V_{G,o})=V_{C} (x_{p} ) 
\end{equation} 

In case B equation (\ref{eq10}) becomes:
 
\begin{eqnarray}
\label{eq13}
R(x)=2\gamma (x)m_{o}c^{2} =2m_{o}c^{2}+2(\gamma (x)-1)m_{o} c^{2} =2m_{o}c^{2} +2K(x) \nonumber\\
=2m_{o}c^{2}+V_{C,o}+V_{G,o}-V_{C}(x)-V_{G}(x)
\end{eqnarray} 
 
and accounting for $-V_{G,o}\approx 10^{-36}\approx 0$ and dividing by $2\gamma_{p}m_{o}c^{2}=m_{p}c^{2}$ one obtains:

\begin{eqnarray}
\label{eq14} 
\gamma (x)/\gamma _{p}=\mu (x)=\rho\left(1-\frac{1}{x} +V_{G} (x)/V_{C,o} \right) \nonumber\\
=\gamma _{p}^{-1} +\rho \left(1-\frac{1}{x} +y(x)\right)=\rho \left(1-\frac{1}{x} +y(x)\right) 
\end{eqnarray}
 
where the last equality holds since, as already noted, $m_{o}\ll m_{p} $, thus $\gamma_{p}\gg1$.For $\rho=1$ equations (\ref{eq11}) and (\ref{eq14}) coincide.

\subsection{The variation of gravitational potential energy with distance}
The gravitational potential energy, $V_{G} (x)$, can be computed using (\ref{eq2}) for any x from:
 
\begin{equation}
\label{eq15} 
-(V_{G}(x)-V_{G,o})=\int _{{\rm \; }x'=1}^{{\rm \; }x'=x}-F_{G}(x'){\rm \; } d(r_{o} x')=\frac{Gm_{o}^{2} }{r_{o} } \int _{{\rm \; }x'=1}^{{\rm \; }x'=x}\frac{\gamma ^{6} (x')}{x'^{2}}dx' 
\end{equation}

where $x'$ denotes the dummy variable.

Using the definition of $y(x)$, it follows:
 
\begin{equation}
\label{eq16} 
y(x)=-(V_{G}(x)-V_{G,o})/V_{C,o}=-(V_{G}(x)-V_{G,o})/Q(e^{2}/\varepsilon r_{o} )=-(V_{G}(x)-V_{G,o})/\rho m_{p}c^{2}  
\end{equation}

and thus: 
 
\begin{equation}
\label{eq17}
y(x)=-\frac{(V_{G}(x)-V_{G,o})}{Q(e^{2}/\varepsilon r_{o})}=\frac{Gm_{o}^{2}}{Q(e^{2} /\varepsilon)}\int_{x'=1}^{x'=x}\frac{\gamma ^{6}(x')}{x'^{2}}dx' 
\end{equation}

and using $m_{p} =2m_{o} \gamma _{p} $ one obtains:

\begin{equation}
\label{eq18} 
y(x)=\frac{\varepsilon Gm_{p}^{2} }{4Qe^{2} } \gamma _{p}^{-2}\int_{x'=1}^{x'=x}\frac{\gamma ^{6} (x')}{x'^{2} }dx'=\frac{\varepsilon Gm_{p}^{2} }{4Qe^{2} } \gamma _{p}^{4}\int_{x'=1}^{x'=x}\frac{\gamma ^{6} (x')/\gamma _{p}^{6} }{x'^{2} }dx' 
\end{equation} 

Using (\ref{eq11}) to express $\gamma (x')$ in (\ref{eq18}) one obtains: 
 
\begin{eqnarray}
\label{eq19}{y(x)=-\frac{(V_{G}(x)-V_{G,o})}{V_{C,o}}=\frac{\varepsilon Gm_{p}^{2} }{4Qe^{2} } \gamma _{p}^{4} \int_{x'=1}^{x'_{p}=x}\frac{\mu ^{6} (x')}{x'^{2}}dx'=} \nonumber\\
=4b\int _{x'=1}^{x'=x}\frac{\mu ^{6} (x')}{x'^{2} }dx'=4b\int _{x'=1}^{x'=x}\frac{\left[1-\rho (1/x'-y(x'))\right]^{6} }{x'^{2} }dx'
\end{eqnarray} 

with:
\begin{eqnarray}
\label{eq20}
4b=\frac{\varepsilon Gm_{p}^{2}}{4Qe^{2}}\gamma_{p}^{4}{\qquad} or{\qquad} 4b=\xi\left(\frac{\gamma_{p}^{4}}{4Q}\right)\nonumber\\
\xi =\frac{\varepsilon Gm_{p}^{2}}{e^{2}}{\qquad};{\qquad}4b=\xi \left[\left(\frac{m_{p}}{2m_{o}}\right)^{4}/4Q\right]
\end{eqnarray} 

The third equality (\ref{eq20}) defines $\xi $, which is the constant we aspire to determine, while the last equality shows that the parameter b is uniquely defined by $\xi $ and by the choice of the mass and charge of the two particles.

 The measurement of $G$ is commonly carried out using metal rod torsion balances \cite{Mohr05,Gundlach00,Gillies97} On the basis of the CODATA recommended \cite{Mohr05} experimental values of $\varepsilon $, G, $m_{p} $ and e (Table 1) it is: 
 
\begin{equation}
\label{eq21} 
\xi_{\exp}=8.09335\cdot 10^{-37}=0.134357(\alpha /2\pi )^{12}
\end{equation} 

Thus if, as an example, one uses this experimental value for  $\xi $ and also chooses $Q=2/9$ (corresponding to $q_{1} /e=2/3$ and $q_{2} /e=1/3$, the charges of $u$ and $\bar{d}$ quarks) and $m_{o} =(1/4)(\alpha /2\pi )^{3} m_{p} =0.3675{\rm \; }eV/c^{2} $, (close to the estimated heaviest neutrino mass of ${0.4\; \; eV/c^{2}} $ \cite{Griffiths08}) then (\ref{eq20}) gives $b=0.6046$, which lies in the range of b values leading to particle confinement as shown below. 

Upon differentiation of (\ref{eq19}) using Leibnitz's rule one obtains:
 
\begin{equation}
\label{eq22} 
\frac{dy(x)}{dx} =4b\frac{\left(1-\rho /x+\rho y(x)\right)^{6}}{x^{2}}\; \;;\; \; with\; B.C.\; \;y(1)=0\; \; ;\; \; case\;\; A 
\end{equation} 

In case B in view of (\ref{eq14}) the differential equation (\ref{eq22}) becomes:

\begin{equation}
\label{eq23} 
\frac{dy(x)}{dx} =4b\rho ^{6} \frac{(1-1/x+y)^{6} }{x^{2}} \; \;;\; \; with\; B.C.\; \;y(1)=0\; \; ;\; \; case\;\; B
\end{equation}

When $\rho =1$ equations (\ref{eq22}) and (\ref{eq23}) are identical.

Once $b$ and $\rho $ have been chosen and the differential equation (\ref{eq22}) or (\ref{eq23}) has been solved numerically to obtain $y(x)$, then recalling (\ref{eq11}) in case A or (\ref{eq14}) in case B and the definition of $y(x)$ it is:

\begin{equation}
\label{eq24} 
\mu (x)=1-\rho \left(\frac{1}{x} -y(x)\right){\qquad};{\qquad} case\;\; A 
\end{equation} 

\begin{equation}
\label{eq25} 
\mu (x)=\rho \left(1-\frac{1}{x} +y(x)\right){\qquad};{\qquad} case\;\; B 
\end{equation}

and thus in either case the function $\mu (x)$ is determined. Thus both the potential energy profiles and also the force profiles are readily determined. Thus from Coulomb's law the Coulombic potential energy is:

\begin{equation}
\label{eq26} 
V_{C} (x)=V_{C,o} /x 
\end{equation}
 
and from the definition of $y(x)=-(V_{G}(x)-V_{G,o})/V_{C,o} $ the gravitational potential energy is:
 
\begin{equation}
\label{eq27} 
-(V_{G}(x)-V_{G,o})=V_{C,o} y(x) 
\end{equation} 

Examples of the $\mu (x)$, $V_{C} (x)$ and $V_{G} (x)$ profiles for $\rho =1$ (thus $K_o=0$ and $-V_{G,o}=\xi V_{C,o}\approx 10^{-36} V_{C,o}=0$) and two different $b$ values ($b=2/9$ and $b=8/15$) are presented in Figures 2 and 3. These two $b$ values have been chosen to lie below and above a critical $b$ value $(b_{c} =0.32=8/25)$  above which, as shown below, particle confinement occurs. 

 As shown in Figure 2a for $b=2/9$ the gravitational potential $V_{G} $ energy approaches a limiting value of $-220{\rm \; MeV}$ for $x\to \infty $ while for $b=8/15$ (Fig. 2b) $-V_{G} (x)$ approaches infinity at a distance of $5\cdot 10^{-3} {\rm \; fm}$ (asymptotic freedom behaviour \cite{Gross73,Politzer73,Cabibbo75}).
\begin{figure}[h]
\centering
\subfigure(a)
\label{fig2:sub:a}
\includegraphics[width=6.50cm,height=6.00cm]{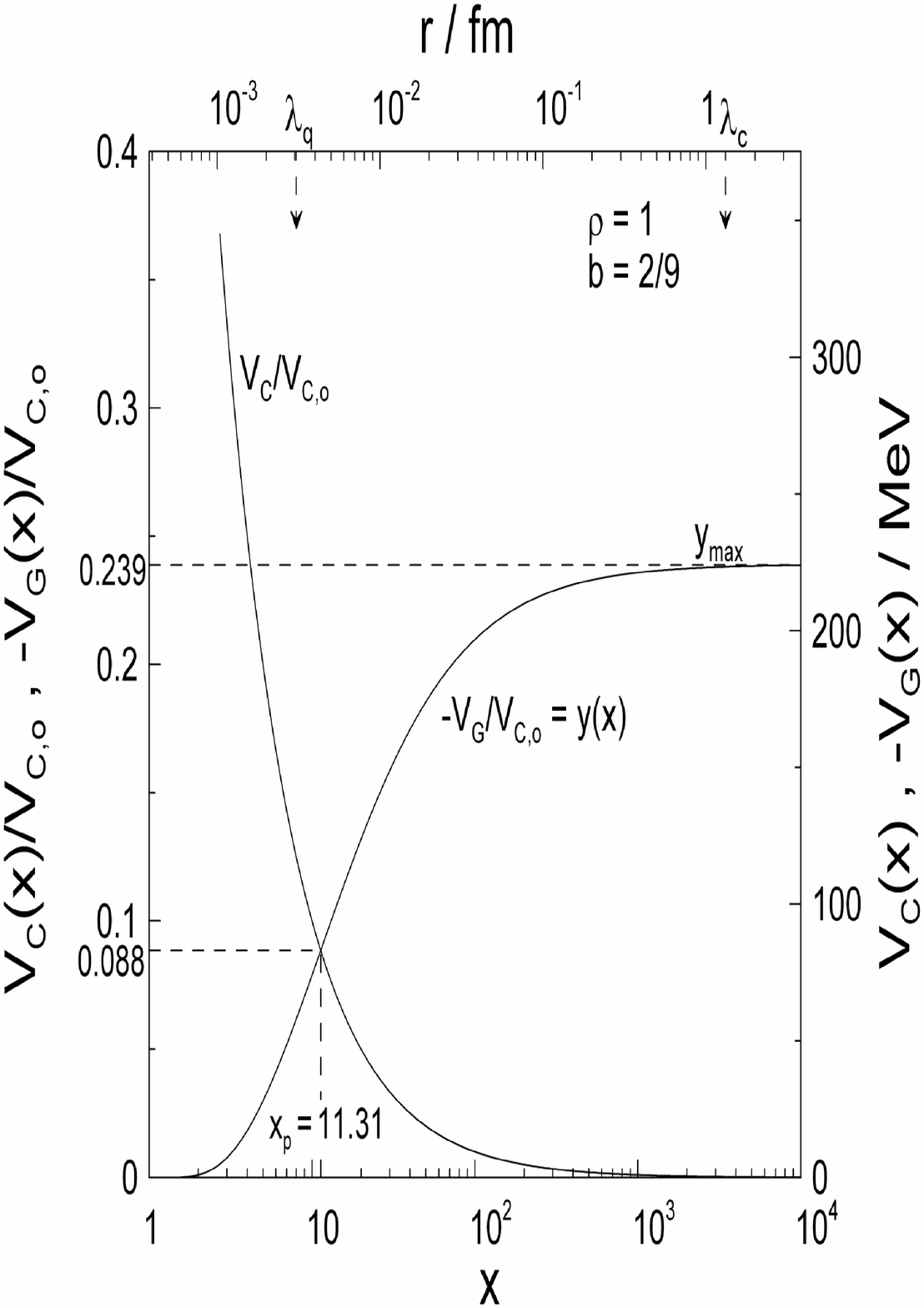}
\hspace{0.0cm}
\centering
\subfigure(b)
\label{fig2:sub:b}
\includegraphics[width=6.50cm,height=6.00cm]{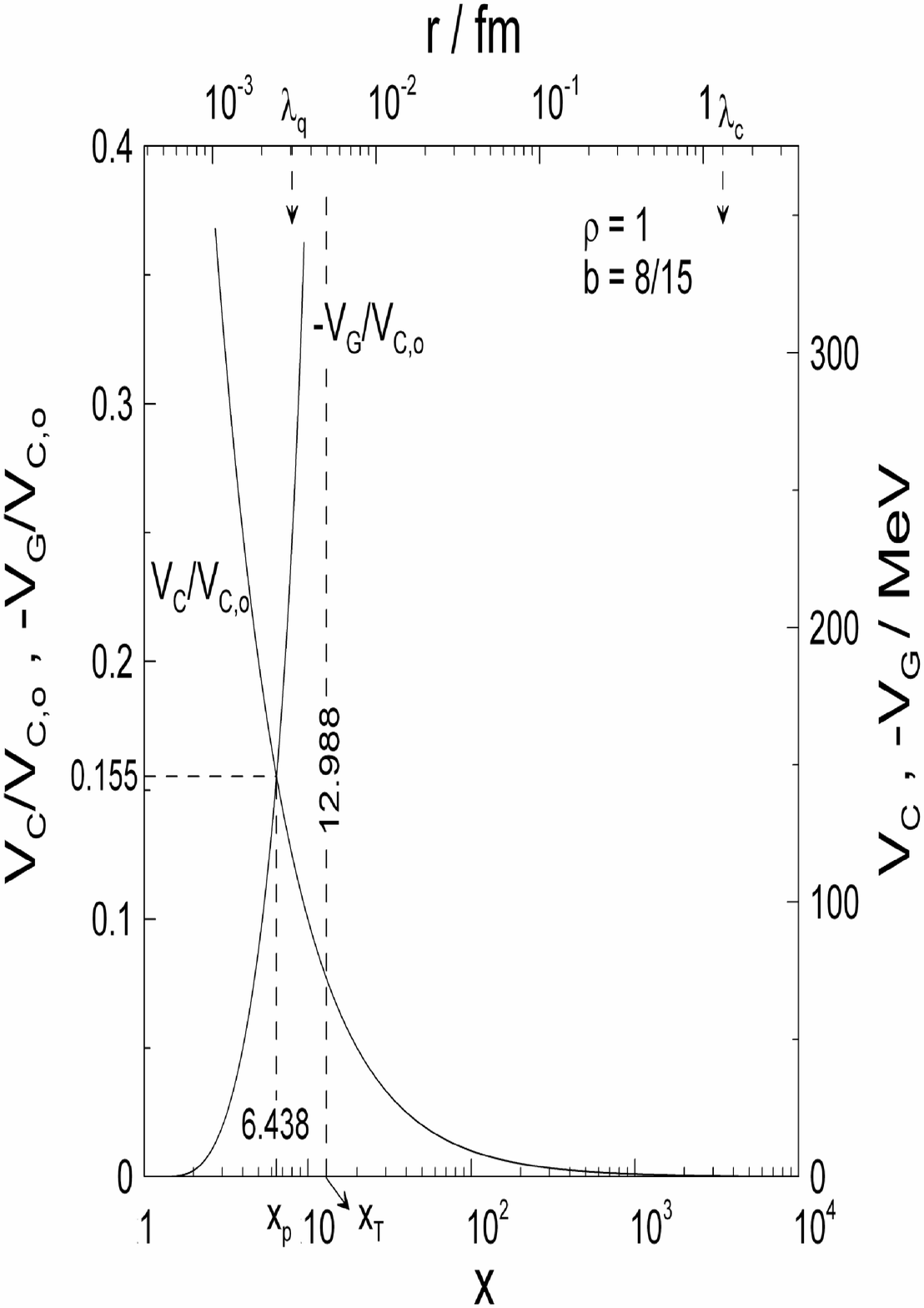}
\caption{Coulombic $(V_{C} )$ and gravitational $(V_{G} )$ potential energy profiles for $\rho =1$, from (\ref{eq26}) and (\ref{eq27}), with $b=2/9$ (a) and $b=8/15$ (b).}
\label{fig2:sub}
\end{figure}
 This bevavior is typical for b values above $b_{c} =8/25$ and leads, as shown below, to particle confinement. Figures 3a and 3b show the corresponding $\mu (x)$ and $R(x)$ profiles for $b=2/9$ and $b=8/15$ respectively. One observes that in the former case $R(x)/c^{2}$ approaches an asymptotic value $(1157{\rm \; MeV/c}^{{\rm 2}} )$, interestingly very close to the masses ($1115$ to $1197{\rm \; MeV/c}^{{\rm 2}} $) of the $\xi $, $\Sigma ^{+} $, $\Sigma ^{o} $ and $\Sigma ^{-} $  spin $1/2$ baryons \cite{Griffiths08}. In the latter case $(b=8/15)$, $\mu (x)$ and $R(x)$ rise steeply after $x_{p} $ (where $\mu (x)=1$ and $R(x)=938.272{\rm \; MeV}$) at some point $x_{T} $ which, as shown below, corresponds to the terminal (maximum) distance during the oscillations of the two particle system. It is again interesting to note (Fig. 3b) that the rest masses of all the spin $3/2$ baryons ($\Omega ^{-} $, $\Xi ^{*} $, $\Sigma ^{*} $ and $\Delta $) \cite{Griffiths08} all correspond to $x$ values between $x_{p} $ and $x_{T} $.
\begin{figure}[h]
\centering
\subfigure(a)
\label{fig3:sub:a}
\includegraphics[width=6.50cm,height=6.00cm]{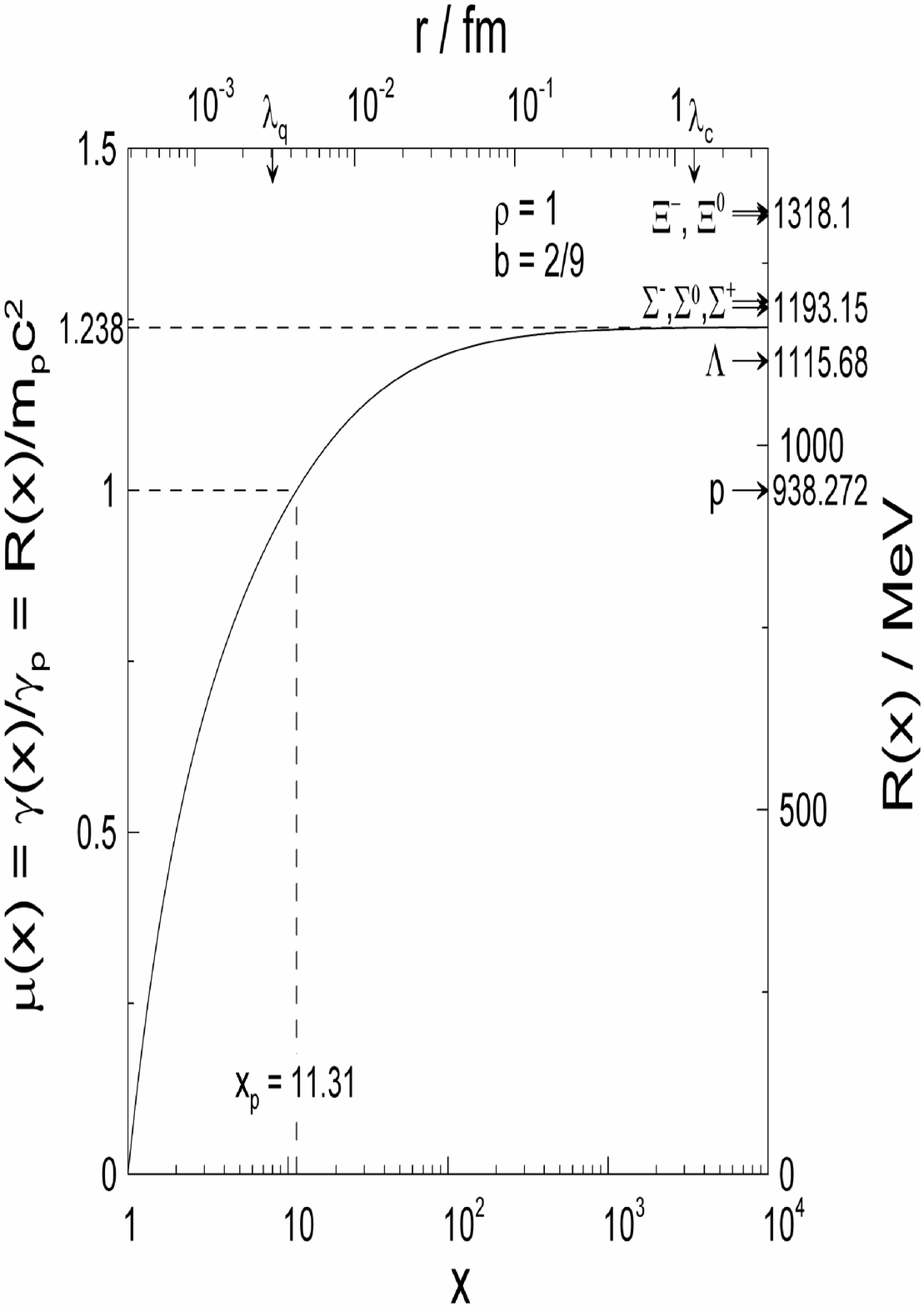}
\hspace{0.0cm}
\centering
\subfigure(b)
\label{fig3:sub:b}
\includegraphics[width=6.50cm,height=6.00cm]{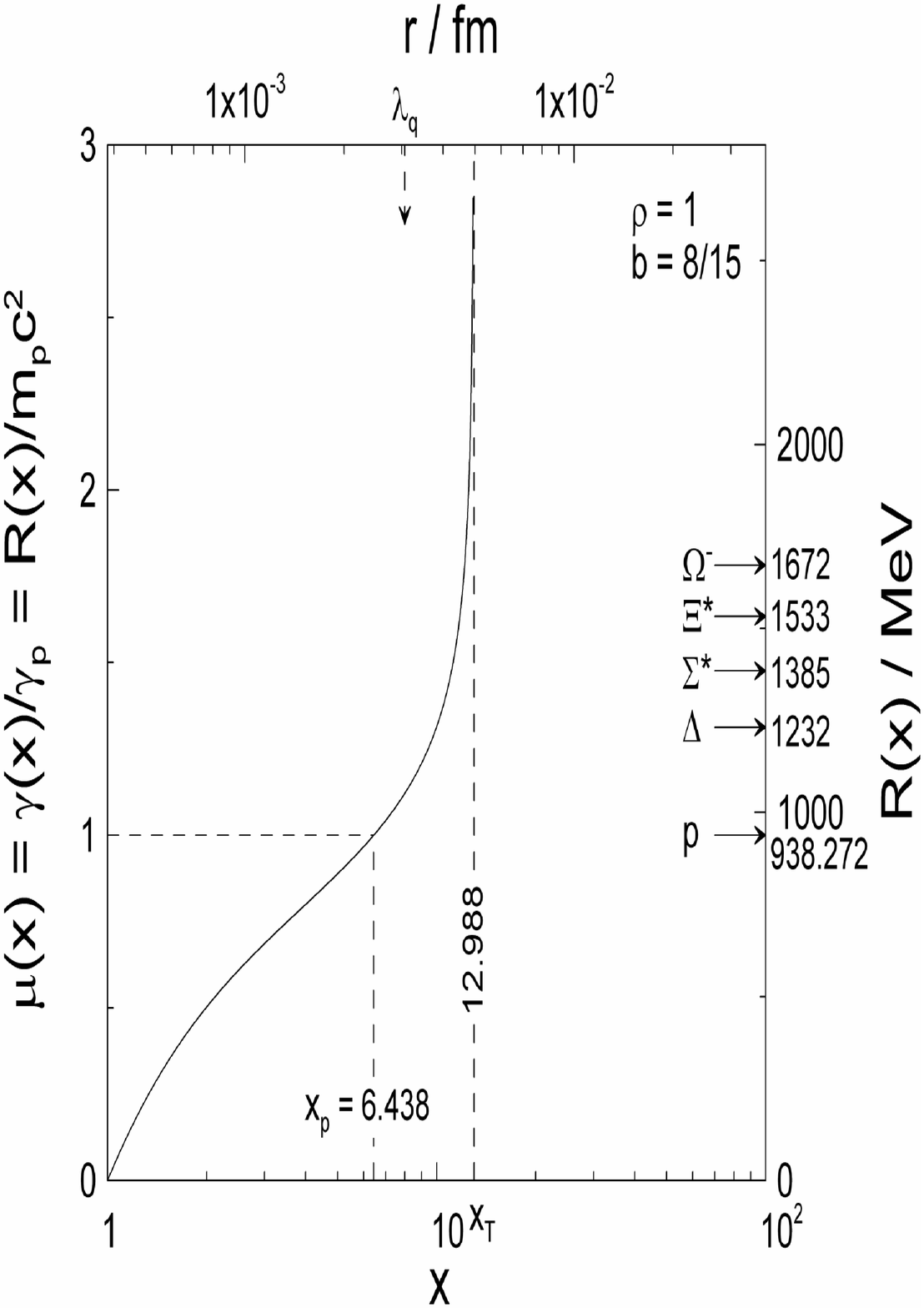}
\caption{Rest mass $R(x)$ and corresponding $\mu (x)$ profiles for $\rho =1$, from (\ref{eq24}) or (\ref{eq25}), with $b=2/9$ (a) and $b=8/15$ (b). Comparison with the rest masses of the spin (1/2) (a) and the spin (3/2) baryons \cite{Griffiths08}.}
\label{fig3:sub}
\end{figure}

\subsection{Force profiles, maximum momentum and confinement conditions}

The force profiles $F_{C} (x)$ and $F_{G} (x)$ can be obtained either via differentiation of  the potential energy profiles (\ref{eq26}) and (\ref{eq27}) or directly from Coulomb's and Newton's Laws, i.e.:
 
\begin{equation}
\label{eq28} 
F_{C} (x)=\frac{Qe^{2} }{\varepsilon r^{2} } =\frac{Qe^{2} }{\varepsilon r_{o} r_{o} x^{2} } =\frac{V_{C,o} }{r_{o} x^{2} } =\frac{2\rho ^{2} m_{p} c^{2} }{Q\lambda _{q} x^{2} }
\end{equation}
 
where in the last equality we have used (\ref{eq3}), (\ref{eq4}) and (\ref{eq5}), and thus also using (\ref{eq20}):
 
\begin{eqnarray}
\label{eq29}
{-F_{G} (x)=\frac{Gm_{o}^{2} \gamma _{\ell }^{2} (x)}{r^{2} } =\frac{Gm_{o}^{2} \gamma ^{6} (x)}{r^{2} } =\frac{Gm_{p}^{2} }{4\gamma _{p}^{2} (Qe^{2} /\varepsilon )} \cdot \frac{(Qe^{2} /\varepsilon )\gamma ^{6} (x)}{r_{o} r_{o} x^{2} } =} \nonumber\\ {=4b\gamma _{p}^{-6} \frac{V_{C,o} \gamma ^{6} (x)}{r_{o} x^{2} } =4b\mu ^{6} (x)F_{C} (x)} 
\end{eqnarray} 
 
Therefore the net force $F(x)$ is given by:
 
\begin{equation} 
\label{eq30} 
F(x)=F_{C} (x)+F_{G} (x)=(1-4b\mu ^{6} (x))F_{C} (x) 
\end{equation}
 
and thus a necessary, but not sufficient, condition for particle confinement is that there exists a distance $x_{m} $, such that: 
 
\begin{equation} 
\label{eq31} 
\mu ^{6} (x_{m} )=1/4b 
\end{equation}
 
At such a point it follows from (\ref{eq1}) and (\ref{eq30}) that the momentum, $p(x)$, exhibits a local maximum.

Although not directly needed here, we note that numerical integration of (\ref{eq22}), i.e. case A, choosing various $\rho $ and $b$ values shows that for all $\rho $ values above 0.7 the condition (\ref{eq31}) is equivalent to:
 
\begin{equation} 
\label{eq32} 
b>b_{\min } =0.136244 
\end{equation}
 
while in case B the maximum momentum condition (\ref{eq32}) is replaced by:
 
\begin{equation} 
\label{eq33} 
b\rho^{6}>b_{\min}=0.136244 
\end{equation} 
 
The force profiles for $\rho=1$ and $b=2/9$ and $8/15$ are given in Figure 4. One observes that in the latter case, asymptotic freedom behavior is obtained \cite{Gross73,Politzer73,Cabibbo75}, i.e. the attractive force $-F_{G} $, which is small at short distances, approaches infinity with increasing distance. Thus in this case ($\rho =1$,$b=8/15$) the two particles cannot escape from each other. 

\begin{figure}[h]
\centering
\subfigure(a)
\label{fig4:sub:a}
\includegraphics[width=6.50cm,height=6.00cm]{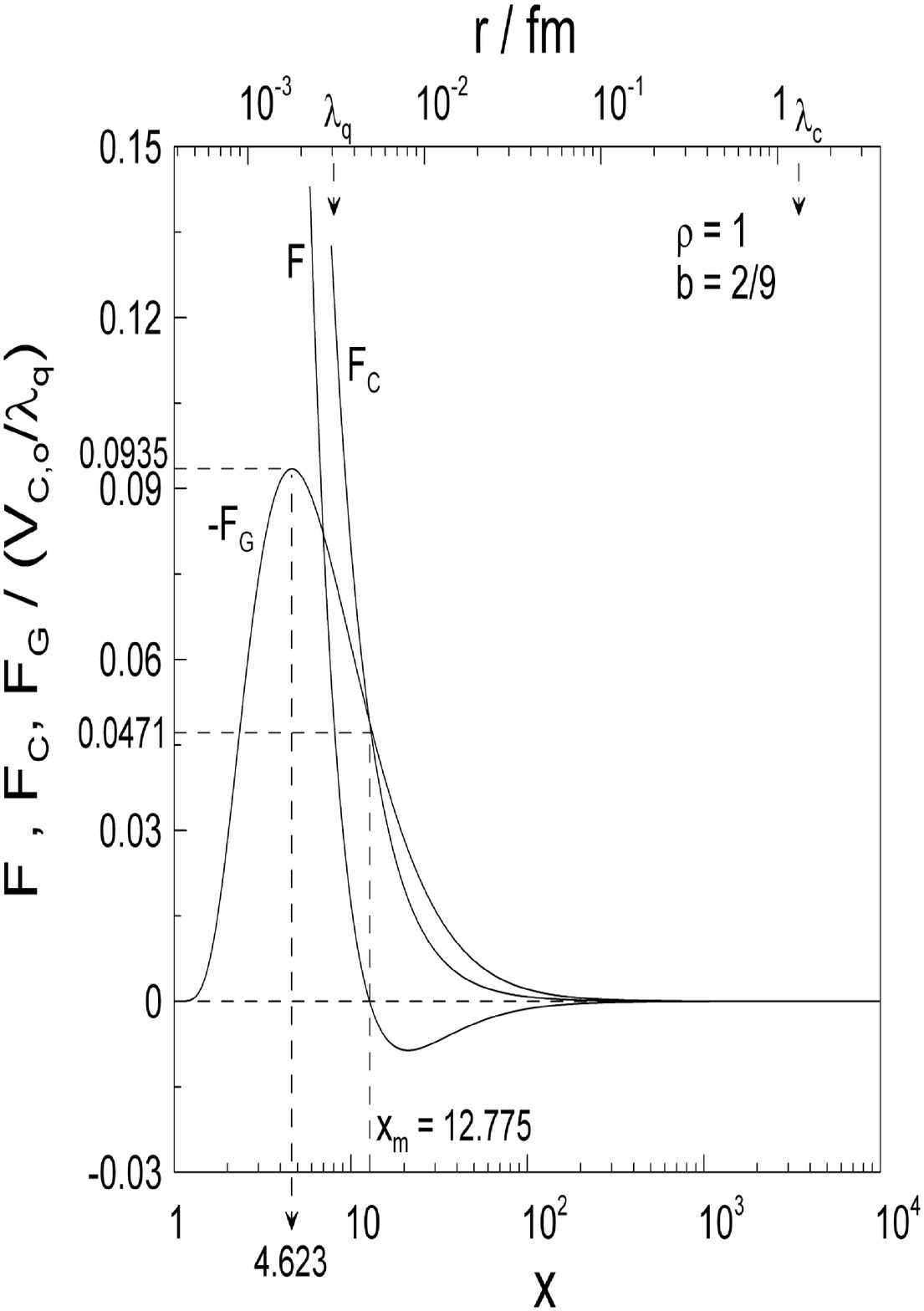}
\hspace{0.0cm}
\centering
\subfigure(b)
\label{fig4:sub:b}
\includegraphics[width=6.50cm,height=6.00cm]{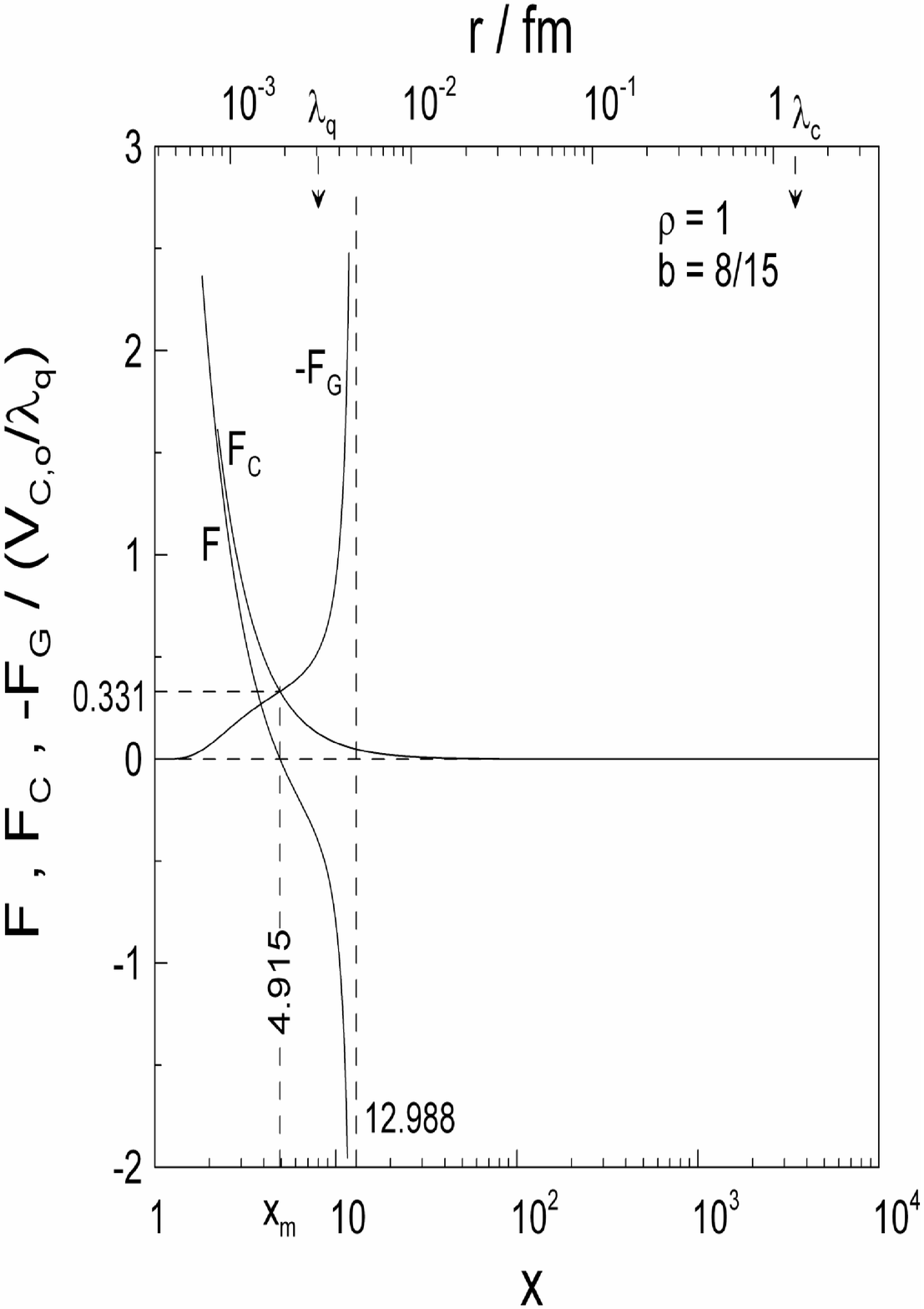}
\caption{Coulombic $(F_{C} )$, gravitational $(-F_{G} )$ and total $(F=F_{C} +F_{G} )$ force profiles, from eqs. (\ref{eq28}) to (\ref{eq30}), for $b=2/9$ (a) and $b=8/15$ (b). In the latter case, asymptotic freedom behaviour \cite{Gross73,Politzer73,Cabibbo75} is obtained.}
\label{fig4:sub}
\end{figure}

\section{Momentum profiles and sufficient confinement condition}

Since the force expressions (\ref{eq28}) to (\ref{eq30}) are valid in both cases A and B, the equation of motion and momentum profiles discussed in this section are the same for both sets of initial conditions A and B. The only difference is that in case B  the function $\mu (x)$ has to be computed from (\ref{eq25}) rather than from (\ref{eq24}). 

A necessary and sufficient condition for particle confinement is obtained by using (\ref{eq2}) and (\ref{eq30}) and examining the equation of motion of particle B:
 
\begin{equation} 
\label{eq34} 
\frac{dp}{dt} =F(x)=F_{C} (x)+F_{G} (x)=F_{C} (x)\left[1-4b\mu ^{6} (x)\right] 
\end{equation}
 
where $p(x)=\gamma(x)m_{o}v(x)$. This equation which can also be written as:
 
\begin{equation}
\label{eq35} 
\frac{dp}{dr} \cdot \frac{dr}{dt} =F(x){\rm \; \; ;\; \; }\frac{dp(x)}{r_{o} dx} v(x)=F(x){\rm \; \; ;\; \; }\frac{dp(x)}{dx} =\frac{\pm \lambda _{q} F_{C} (x)}{8\rho c\left(1-\gamma _{(x)}^{-2} \right)^{1/2} } \left[1-4b\mu ^{6} (x)\right] 
\end{equation}
 
where we have used (\ref{eq5}) with $Q=1/4$ (identical particles), thus, $r_{o} =(1/8\rho)\lambda_{q} $, and have also used the definition of the Lorentz factor , $\gamma $ i.e. $(v/c)=\pm(1-\gamma^{-2})^{1/2}$. Using (\ref{eq28}) to express $F_{C}(x)$ one obtains:
 
\begin{equation}
\label{eq36} 
\frac{\left|dp(x)\right|}{dx} =\frac{\rho m_{p}c}{\left(1-\gamma^{-2}(x)\right)^{1/2}x^{2}}\left[1-4b\mu^{6}(x)\right]   
\end{equation} 
 
Integration of this equation gives the momentum profile $p(x)$, i.e.:
 
\begin{equation} 
\label{eq37} 
\left|p(x)\right|=p_{o}+\rho m_{p} c\int_{x'=1}^{x'=x}\frac{1-4b\mu^{6} (x')}{\left(1-\gamma^{-2}(x')\right)^{1/2} x'^{2}}dx'
\end{equation}
 
where $p_{o}$ is the momentum at $x=1$. This equals zero in case B $(K_{o}=0)$ and can be easily shown from ($\ref{eq9}$) to equal, to a very good approximation, $m_{p}c(1-\rho)$ in case A. Thus using also (\ref{eq18}) to express $\gamma _{p} (=(4b/\xi )^{1/4} )$ one obtains:
 
\begin{eqnarray}
\label{eq38}
\nonumber \frac{\left|p(x)\right|}{(m_{p} /2)c}=2(1-\rho )+2\rho \int _{{\rm \; }x'=1}^{{\rm \; }x'=x}\frac{1-4b\mu ^{6} (x')}{\left(1-\gamma ^{-2} (x')\right)^{1/2} x'^{2} }dx'{\qquad};{\qquad} case\; A\\ \frac{\left|p(x)\right|}{(m_{p} /2)c} =2\rho \int _{{\rm \; }x'=1}^{{\rm \; }x'=x}\frac{1-4b\mu ^{6} (x')}{\left(1-\gamma ^{-2} (x')\right)^{1/2} x'^{2}}dx'{\qquad\qquad\qquad};{\qquad} case\; B  
\end{eqnarray}
 
Thus, focusing on case B, a necessary and sufficient condition for particle confinement and establishment of a self-sustained oscillation is that there exists some finite positive distance $x_{T} $ at which $p(x_{T} )=0$, i.e. the momentum vanishes. At this point it is $dp(x)/dt<0$ and thus at $x_{T} $ the velocity becomes negative until $x=1$ is reached thus completing the cycle (Figure 5).
\begin{figure}
\centerline{\includegraphics[width=8.75cm,height=8.74cm]{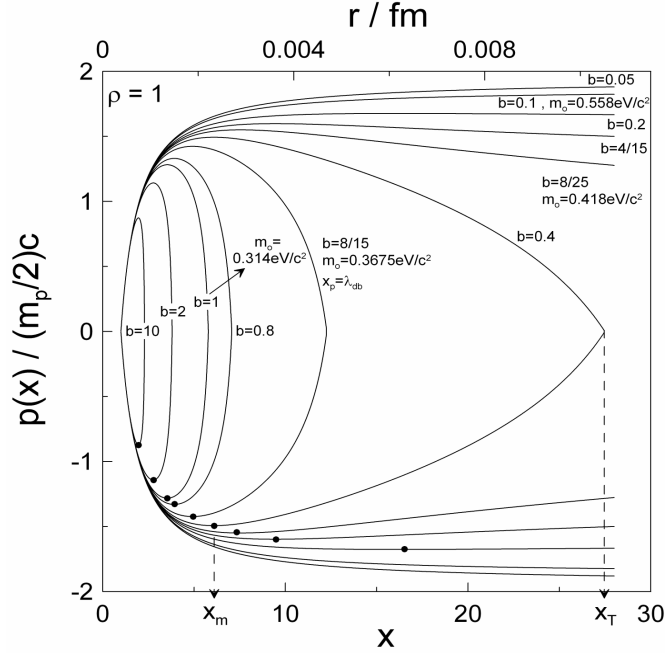}}
\caption{Particle momentum profiles from (\ref{eq38}) for $\rho =1$ and various values of b and corresponding rest particle mass $m_{o} $, also showing the definitions of $x_{m} $and $x_{T} $ for the case $b=0.4$. The momentum exhibits a local maximum and minimum for $b>b_{\min}$$(m<m_{o,\max}=0.516{\rm \; eV/c}^{2})$ and the particles are confined (finite $x_{T} $) for $b>b_{c} =8/25$ $(m<m_{o,c} =0.418{\rm \; eV/c}^{2} )$. It is $b=(\epsilon Gm^2_p/16Qe^2)(m_p/2m_o)^4$}
\label{fig5}
\end{figure}

In this case the period of the oscillation, $T_{osc} $, can be computed using (\ref{eq34}), i.e. 
 
\begin{equation} 
\label{eq39} 
dt=\frac{dp(x)}{F(x)}=\frac{(dp(x)/dx)}{F(x)}dx{\qquad};{\qquad}(T_{osc} /2)=\int_{x'=1}^{x'=x_{T}}\frac{(dp(x')/dx')}{F(x')}dx'
\end{equation}
 
e.g. for case B, using (\ref{eq35}):
 
\begin{equation} 
\label{eq40} 
{\rm T}_{{\rm osc}} /2=\frac{\lambda _{q} }{8\rho c}\int_{x'=1}^{x'=x_{T}}\frac{dx'}{(1-\gamma ^{-2}(x'))^{1/2}}
\end{equation} 
 
 It should be noted that the profile $\gamma (x)$ depends on $\rho $ and b, but since $\gamma (x)\gg 1$ except in the vicinity $x\approx 1$, one can easily show via integration for practically any $\rho $ and $b$ values that to a very good approximation  the integral equals $(x_{T} -1)$ and thus:
 
\begin{equation} 
\label{eq41} 
T_{osc}=(2.55\cdot 10^{-27}s)\; \rho^{-1}\int_{x'=1}^{x'=x_{T}}\frac{dx'}{(1-\gamma ^{-2} (x'))^{1/2}}=(2.55\cdot10^{-27}\rm s)\; \rho ^{-1}(x_{T}-1)
\end{equation}
 
where $x_{T}$ is the maximum (terminal) value of x during an oscillation (Fig. 5).

Figure 5 shows the momentum profiles obtained for $\rho =1$ (thus applicable to both cases A and B) and various values of b $(=\xi (m_{p} /2m_{o} )^{4} /16Q)$. Thus increasing b corresponds to decreasing $m_{o} $ and the value $b=8/15=0.5333$ corresponds, using $\xi _{\exp } $ (Table 1), to $m_{o} =0.3675{\rm \; eV/c}^{{\rm 2}} $. The $p(x)$ profile exhibits a maximum for all $b>b_{\min } =0.136244$ but particle confinement occurs only for $b>b_{c} =0.32=8/25$, i.e. for $m_{o} <0.418{\rm \; eV/c}^{{\rm 2}} $. This is shown more clearly in Figure 6a, based on Fig. 5, which shows the dependence of $x_{m} $ (distance of maximum momentum), of $x_{T} $ (terminal, i.e. maximum distance during an oscillation) and of $x_{p} $ (distance of $\mu (x_{p} )=1$) on b and thus $m_{o}$.

\begin{figure}[h]
\centering
\subfigure(a)
\label{fig6:sub:a}
\includegraphics[width=5.50cm,height=6.00cm]{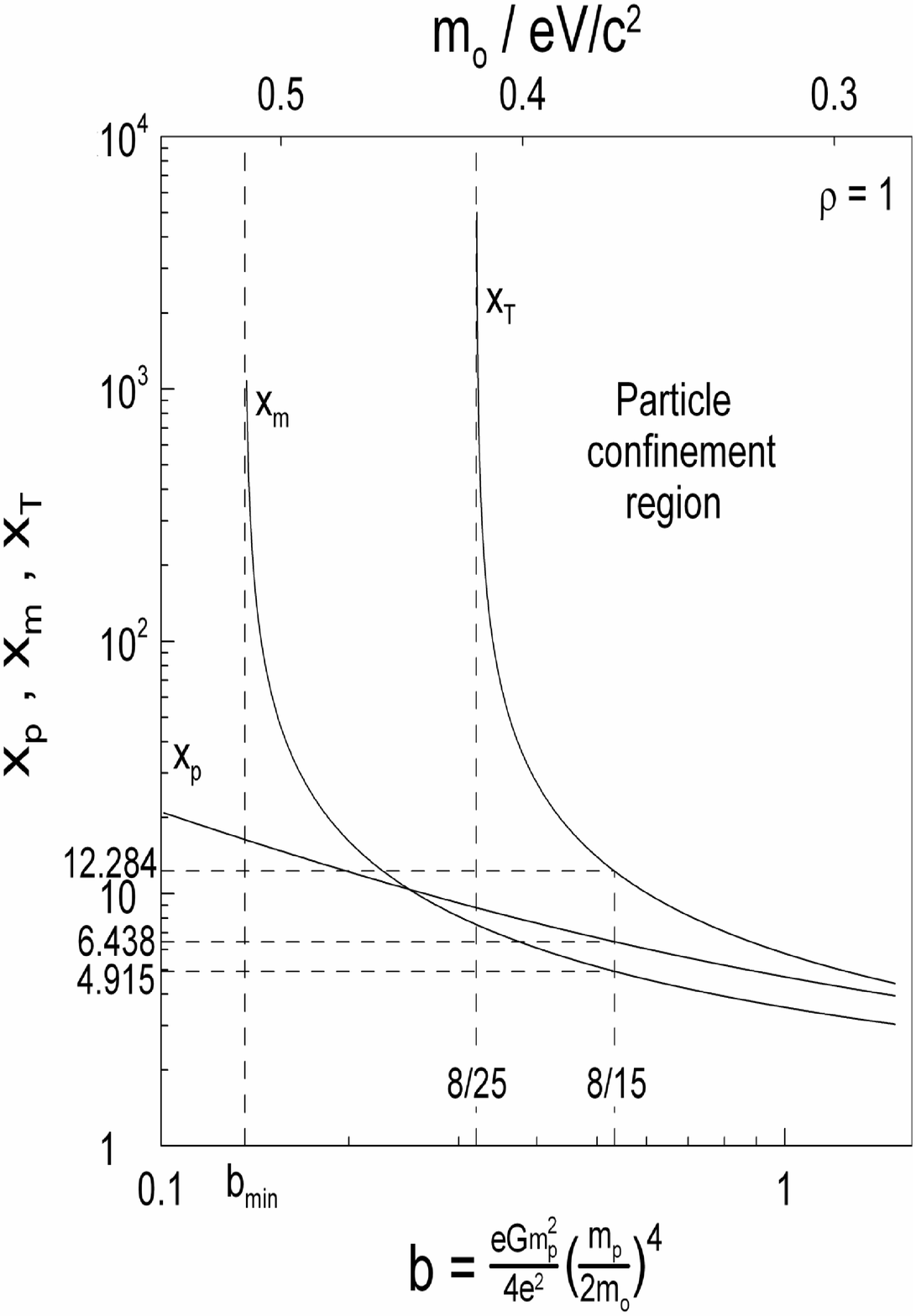}
\hspace{0.0cm}
\centering
\subfigure(b)
\label{fig6:sub:b}
\includegraphics[width=6.50cm,height=6.00cm]{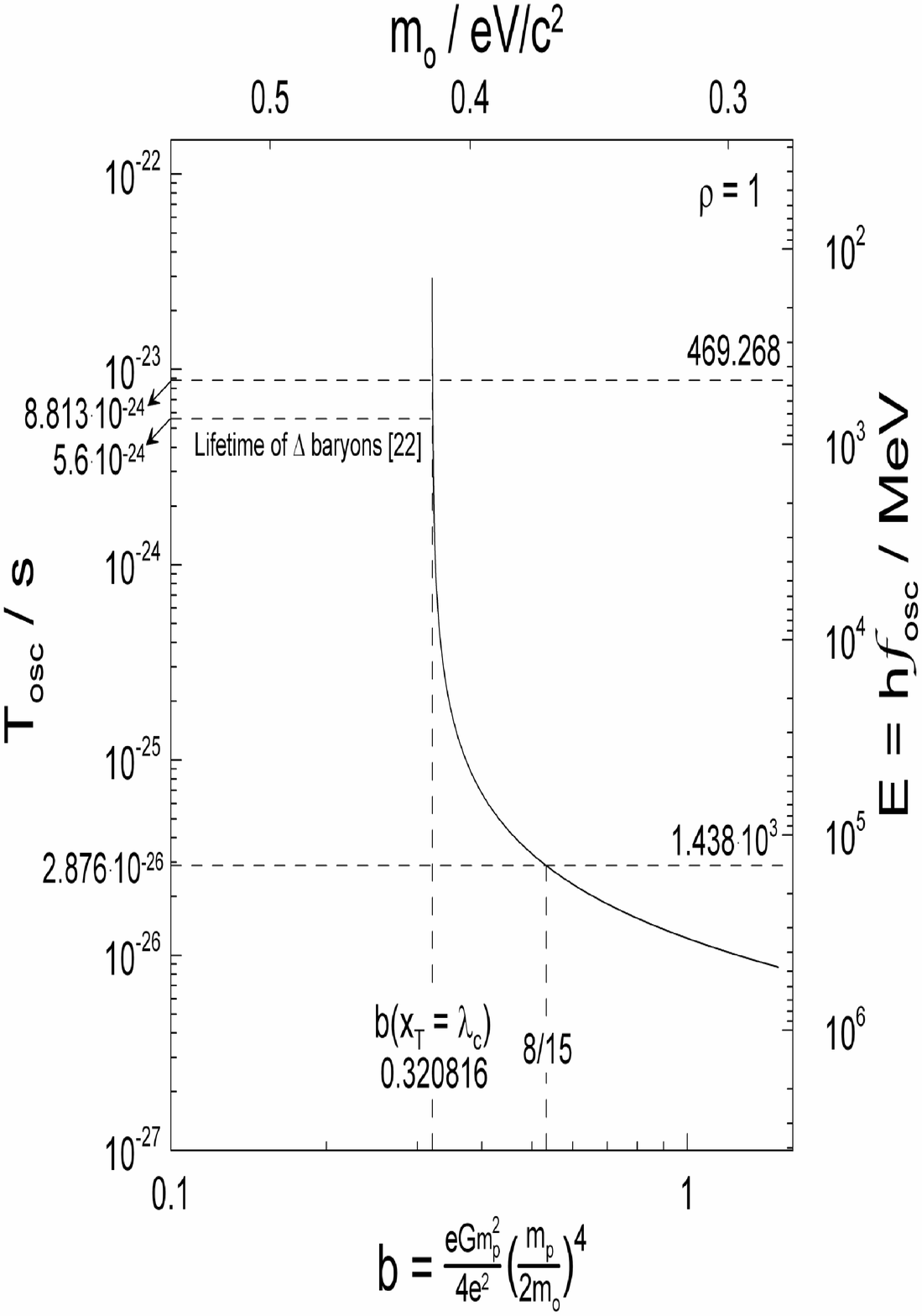}
\caption{Effect of $b$ or of rest mass $m_{o} $ on $x_{p} $ $(\mu (x_{p} )=1)$, $x_{m} $$(-F_{G} (x_{m} )=F_{C} (x_{m} ))$ and $x_{T} $ (terminal distance during oscillations) for $\rho =1$ (a).Effect of $b$ or of rest mass $m_{o} $ on the period of the oscillator. Comparison of the oscillator period, when $x_{T} $ corresponds to the proton Compton length $\lambda _{c} $, with the lifetime of $\Delta $ baryons \cite{Griffiths08} (b).}
\label{fig6:sub}
\end{figure}
Figure 6b, based on (\ref{eq40}), shows the dependence of the oscillation period $T_{osc} $ on b, thus $m_{o} $. Interestingly $T_{osc} $ in the vicinity of the $b_{c} =0.32=8/25$ limit (i.e. at the b value of 0.320816 corresponding to $x_{T}=r_{o} \lambda _{c} $) equals $8.8\cdot 10^{-24} {\rm \; s}$, which is very close to the lifetime $(5.6\cdot 10^{-24} \rm \; s)$of $\Delta $ baryons \cite{Griffiths08}. 

\subsection{Maximum and critical particle rest mass}
The $x_{p} $ vs b curve of Figure 6a for $\rho =1$ forms the basis, upon axis rotation, of figure 7 by establishing the curve $b(x_{p} )$. Before presenting this figure (Fig. 7) in detail, it is useful to make some observations.

\begin{figure}
\centerline{\includegraphics[width=10.25cm,height=8.74cm]{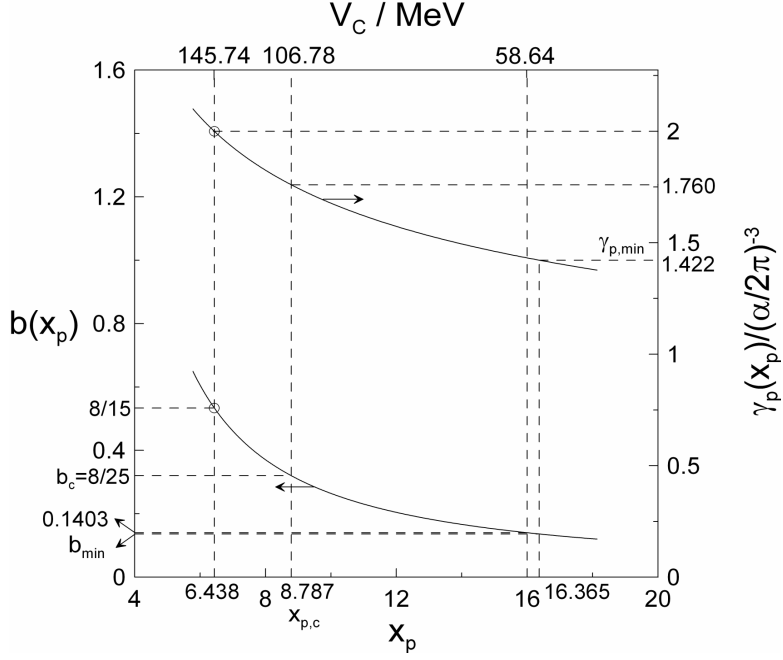}}
\caption{Effect of $x_{p} $ (distance where $\mu (x_{p})=1$, i.e. $R=m_{p} c^{2} $) and corresponding Coulombic potential energy $V_{c} (x_{p})$ for $\rho =1$ on $b(=(eGm_{p}^{2}/4e^{2})(m_{p}/2m_{o})^{4})$ and on $\gamma _{p} (=m_{p} /2m_{o} )$ showing, by the two circled points, the b and $\gamma _{p} $ values corresponding to $m_{o} =(1/4)(\alpha /2\pi )^{3} m_{p} =0.3765{\rm \; eV/c}^{2}$. The choice of any $m_o$ and thus $b(x_p)$ leads via this figure and $\xi=4b/\gamma^4_p$ to $\xi=(2/15)(\alpha/2\pi)^{12}$ (eq. (\ref{eq50})).}
\label{fig7}
\end{figure}

\begin{figure}
\centerline{\includegraphics[width=11.75cm,height=8.50cm]{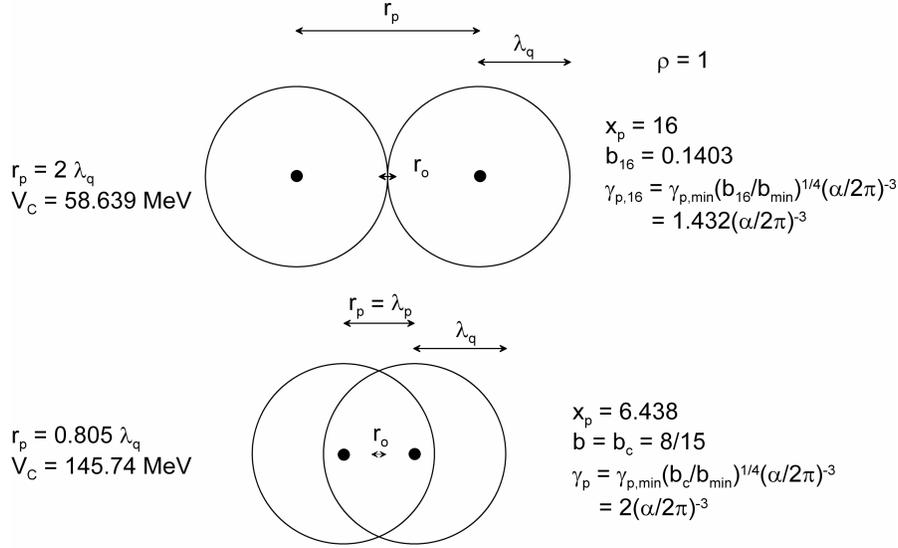}}
\caption{Schematic for $\rho =1$ of the location and classical radii $\lambda_{q} (=3.06\cdot 10^{-3}$\;fm) of the two particles at $x_{p}$ when $b=0.14028$ (thus from Fig. 7, $x_{p} =16$) and when $b=8/15$ (thus $x_{p} =6.4378$). In the latter case it is $\gamma _{p} =2(\alpha /2\pi )^{-3}$.}
\label{fig8}
\end{figure}
In view of the definition of b (Eq. 20), the existence of $b_{\min }$, as a limiting b value for momentum maximization ($(-F_{G}(x_{m})=F_{C}(x_{m}))$ (\ref{eq32}), implies directly the existence of a minimum $\gamma_{p} $ value necessary for momentum maximization, i.e. from (\ref{eq20}) for $Q=1/4$:
\begin{equation} 
\label{eq42} 
{\rm \; }\gamma _{p,\min } ={\rm (}4b_{\min } /\xi )^{{\rm 1/4}} {\rm \; \; \; ;\; \; \; }\xi =\frac{\varepsilon Gm_{p}^{2} }{e^{2}}
\end{equation} 
 
Since from (\ref{eq7}) it is $m_{o} =m_{p} /2\gamma _{p} $, it therefore follows, as already known from Fig. 6a, that there exists a maximum particle, mass, $m_{o,\max } $, above which the momentum exhibits no local maximum and thus particle confinement is not possible. 

The value of $m_{o,\max } $ is readily obtained from: 
 
\begin{equation} 
\label{eq43}  
{\rm \; m}_{{\rm o,max}} =\frac{m_{p} }{2\gamma _{p,\min } } =\frac{m_{p} }{2(4b_{\min } /\xi )^{1/4} }  
\end{equation} 
 
Similarly the existence of $b_{c}=8/25$ as a limiting value for particle confinement, implies the existence of a critical $m_{o} $ value, denoted $m_{o,c} $, above which particle confinement is not possible (Fig. 6a). The value of $m_{o,c} $ it readily obtained from:
 
\begin{equation}
\label{eq44}  
m_{o,c} =\frac{m_{p} }{2(4b_{c} /\xi )^{1/4} }  
\end{equation}

\section{Computation of G}
So far we have not used the experimental G value, which we denote $G_{\exp } =6.674\cdot 10^{-11} {\rm \; m}^{{\rm 3}} kg^{-1} s^{-2} $ \cite{Mohr05,Gundlach00,Gillies97}. Using this value it is $\xi_{\exp}=0.1343595(\alpha /2\pi )^{12} $ (Table 1). 

Substituting in (\ref{eq44}) one finds:
 
\begin{equation} 
\label{eq45}  
\frac{2m_{o,c} }{m_{p}}=\gamma _{p,c}^{-1} =\left(\frac{\xi _{\exp }}{4b_{c} } \right)^{1/4} =\left[\frac{0.1343595(\alpha /2\pi )^{12} }{4(0.32)} \right]^{1/4} =0.5692(\alpha /2\pi )^{3}  
\end{equation} 
 
\begin{equation} 
\label{eq46}
m_{o,c}=0.2846(\alpha/2\pi)^{3}m_{p}{\rm \; \; \; }=0.4183{\rm \; \; eV/c}^{{\rm 2}}  
\end{equation} 
 
\begin{equation}
\label{eq47}
\gamma_{p,c} =1.7568(\alpha /2\pi )^{-3} {\rm \; \; \; \; ;\; \; }b=b_{c} =0.32=8/25 
\end{equation} 
 
 The computed $m_{o,c}$ value lies very close to the estimated heaviest neutrino mass of $0.4{\rm \; eV/c}^{{\rm 2}} $ \cite{Griffiths08} and this is quite encouraging, although neutrinos are neutral. On the other hand (Fig. 7) the $x_{p} $ value, denoted $x_{p,c} $, corresponding to $b_{c} $ is $8.787$ and the Coulombic energy $V_{c} $ is $106.78$ MeV, quite low in comparison to the QCD transition energy of $E_{QCD} =151(6)$ MeV. One then notes that since from (\ref{eq20}) it is: 
 
\begin{equation}
\label{eq48}
\frac{\gamma_{p}(x_{p})}{\gamma _{p} (x_{p,c} )} =\left(\frac{b(x_{p})}{b_{c}}\right)^{1/4}
\end{equation}
 
for $\gamma _{p} (x_{p} )=2(\alpha /2\pi )^{-3} $ it is $b(x_{p} )=0.5333=8/15$ which corresponds to $x_{p} =6.438$, thus (top axis in Fig. 7) $V_{C} =145.74{\rm \; MeV}$, which lies within the current uncertainty limits of $E_{QCD}$ \cite{Aoki06}. It is also quite close to the rest energy (139.6 $MeV$) of $\pi^+$ mesons \cite{Griffiths08,Nambu84, Povh06} which are stable particles and follow electromagnetic decay 

 These observations suggest a clear strategy for selecting the appropriate $m_{o} $ value below $m_{o,c} $.

 The energy of $E_{QCD}(=151(6)$ MeV) of gluons at the transition temperature of QCD \cite{Braun07,Aoki06, Fodor04} defines the minimum allowed distance, $x_{p} $, of the two particles in view of their Coulombic repulsion, i.e. 
 
\begin{equation} 
\label{eq49}
V_{C,o} /x_{p} =E_{QCD} {\rm \; \; ;\; \; }x_{p} =\rho m_{p} c^{2} /E_{QCD}  
\end{equation} 
 
\noindent which for $\rho =1$ and $E_{QCD} =145.74{\rm \; MeV}$ gives $x_{p} =6.438$. This $x_{p} $ value corresponds to $b=0.53333=8/15$ (Fig. 7). In view of $x=8\rho (r/\lambda _{q} )$ (eq. (5)) , this for $\rho =1$ gives $(r/\lambda _{q} )=0.8047$ (Fig. 8), which implies, as is reasonable to expect since a bound state is formed, a significant overlap between the classical radii of the two particles ($r_{p} /\lambda _{q} =2$ corresponds to no overlap, Fig. 8). 

 We thus select $m_{o} =(1/4)(\alpha /2\pi )^{3} m_{p} =0.3675{\rm \; eV/c}^{{\rm 2}} $, which is also very close to the heaviest neutrino mass of $0.4{\rm \; eV/c}^{{\rm 2}}$ \cite{Griffiths08}. Since $2m_{o}\gamma_{p} =m_{p} $, this implies $\gamma _{p} =2(\alpha /2\pi )^{-3} $ as can also be seen in Fig. 7, where the $\gamma_{p}(x_p)$ curve has been constructed from this equality and (\ref{eq48}).

 Thus from the definition of $b(=\xi \gamma _{p}^{4}/4)$ one obtains:
 
\begin{equation}
\label{eq50}
\xi=\frac{\varepsilon Gm_{p}^{2} }{e^{2} } =\frac{4b}{\gamma _{p}^{4}}=\frac{4(8/15)}{2^{4}(\alpha/2\pi)^{-12}} =\frac{2}{15} (\alpha/2\pi)^{12}
\end{equation}
 
It is worth noting that in view of (\ref{eq48}) and $\xi=4b/\gamma^4_p$, the same $\xi$ value is obtained by choosing any point $x_p$ in Figure 7.
Therefore from (\ref{eq50}):
 
\begin{eqnarray} 
\label{eq51}
G=\frac{2}{15} (\alpha /2\pi )^{12} \frac{e^{2} }{\varepsilon m_{p}^{2}}=(2/15)(6.02370395)10^{-36}(0.82464)10^{26}= \nonumber\\=6.62318\cdot 10^{-11}m^{3}/kg^{-1}s^{-2}\end{eqnarray}

\noindent{which is the expression derived in \cite{Vayenas07, Vayenas2007} and is in excellent agreement with the experimental value of $6.674\cdot 10^{-11} {\rm \; \; }m^{3}kg^{-1}s^{-2}$ \cite{Mohr05,Gundlach00,Gillies97}.}

\begin{figure}
\centerline{\includegraphics[width=9.75cm,height=12.74cm]{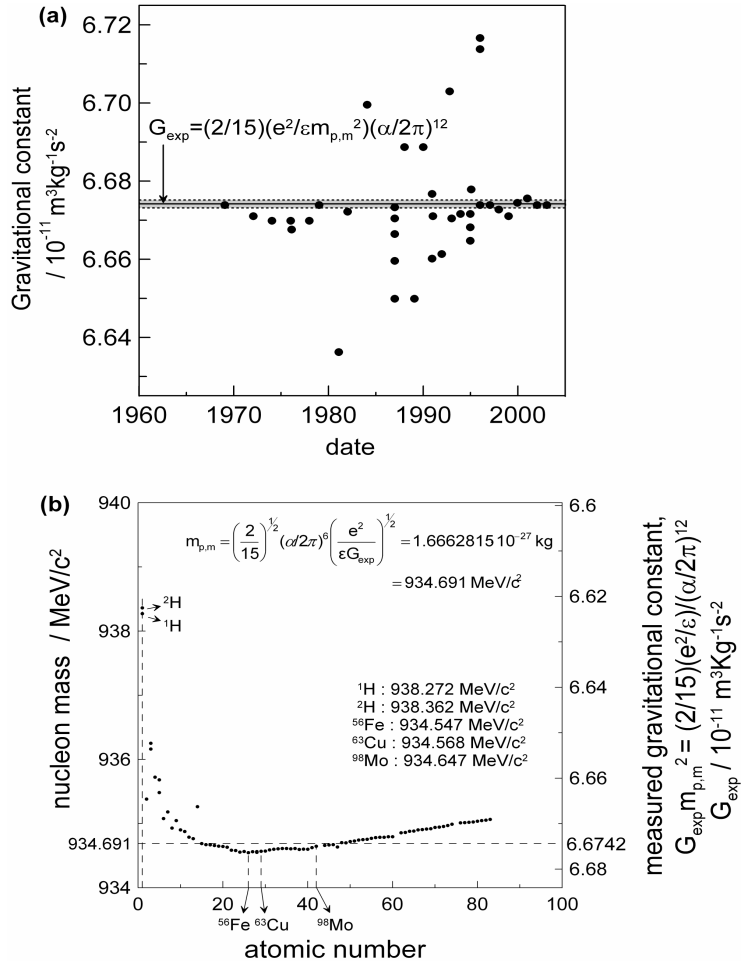}}
\caption{Predicted and measured gravitational constant and average nucleon mass.\textbf{ }(a) Comparison of (\ref{eq51}) with the time evolution of the experimental gravitational constant value, and with some of the most recent experimental values [19,21]; Pre-1997 values data from Table 2 in Ref. 21. Post-1997 values from Table X in Ref. 19. Shaded areas shows the CODATA recommended [19] value of $(6.6742\pm 0.001)\times 10^{-11} {\rm \; }m^{3} kg^{-1} s^{-2} $ (b) Comparison of the average nucleon mass in various nuclei with the one predicted by equation (\ref{eq52}). The right ordinate shows the value of $G_{\exp} $ expected to be measured with different torsion balance rod materials.}
\label{fig9}
\end{figure}

 The agreement becomes quantitative if one accounts for the fact that practically all measurement of G are carried out using torsion balances which utilize metal rods consisting of heavy elements such as Fe or Mo \cite{Mohr05,Gundlach00,Gillies97} rather than protons. Thus using the average nucleon mass in $^{56}Fe$, i.e. $m_{p,m}=1.666157\cdot 10^{-27}{\rm \; kg}$$=934.547{\rm \; MeV/c}^{{\rm 2}} $ or that of  $^{98} Mo(934.647{\rm \; MeV/c}^{{\rm 2}} )$ rather than the proton mass $m_{p} =1.67262171\cdot 10^{-27} {\rm \; kg}$ $=938.272{\rm \; MeV/c}^{{\rm 2}} $, one obtains:
 
\begin{eqnarray} 
\label{eq52}
G_{\exp}=G(m_{p}/m_{p,m})^{2}=6.6760\cdot 10^{-11}\;{m}^3kg^{-1}s^{-2} {\qquad(Fe)}
\nonumber\\{\qquad}=6.6746\cdot 10^{-11} \;{m}^3kg^{-1}s^{-2} {\qquad(Mo)}
\end{eqnarray}
 
which is in quantitative agreement with experiment (Fig. 9a). 
 
Conversely when solving (\ref{eq51}) for $m_{p,m} $ one obtains:
 
\begin{equation}
\label{eq53} 
m_{p,m} =\left(\frac{2}{15} \right)^{1/2} (\alpha /2\pi )^{6} \left(\frac{e^{2} }{\varepsilon G} \right)^{1/2} =1.666\cdot 10^{-27} {\rm \; kg}=934.6{\rm \; MeV} 
\end{equation} 
 
which is also in excellent agreement with experiment (Fig. 9b) particularly for those elements (Fe, Mo, Cu) commonly used in measuring $G_{\exp} $ with metal rod torsion balances \cite{Mohr05,Gundlach00,Gillies97}. As Fig. 9b shows (\ref{eq52}) predicts a small $(\sim 0.01\%)$ variation in the $G$ value measured with different metal rod torsion balance elements and such a small variation has indeed been observed for years \cite{Mohr05,Gillies97}. A prediction of (\ref{eq52}), (\ref{eq53}) and Figure 9b is that if one could somehow measure $G_{\exp}$ using protons or $H_2$ instead of metal rods, then the measured value would be $6.62318\cdot 10^{-11} {\rm \; m}^{{\rm 3}}kg^{-1} s^{-2}$, i.e. 0.8\% smaller than the experimental value, but this is obviously a very difficult experiment.

\subsection{Maximum $m_{o}$ value}
 From (\ref{eq43}) and the computed value of $\xi$ (eq. 50) one can compute the value of $m_{o,\max}$. Thus: 
 
\begin{equation}
\label{eq54} 
m_{o,\max } =\frac{m_{p} }{2\gamma_{p,\min}}=\frac{m_{p} }{2(4b_{\min } /\xi )^{1/4} } =0.351(\alpha /2\pi)^{3}m_{p} =0.516{\rm \; eV/c}^{{\rm 2}}
\end{equation}
 
Similarly one can use the critical value of b for particle confinement, i.e. $b_{c} =0.32$ to compute the corresponding critical particle mass $m_{o,c} $ above which particle confinement does not occur:
 
\begin{equation}
\label{eq55} 
m_{o,c} =\frac{m_{p} }{2\gamma _{p,c} } =\frac{m_{p} }{2(4b_{c} /\xi )^{1/4} } =0.2814(\alpha /2\pi )^{3} m_{p} =0.4183{\rm \; eV/c}^{{\rm 2}}  
\end{equation} 
 
As already noted, the value of $m_{o} $ corresponding to $\gamma_{p}=2(\alpha /2\pi )^{-3} $  is $0.3675{\rm \; eV/c}^{{\rm 2}}$, i.e. is smaller than both $m_{o,\max } $ and $m_{o,c} $. All these values are in excellent agreement with the estimated heaviest neutrino mass of $0.4{\rm \; eV/c}^{{\rm 2}}$ \cite{Griffiths08}. 

 In brief, although all $m_{o} $ values below $m_{o,c} $ lead to the formation of confined states (Figures 5 and 6) and via (\ref{eq50}) and (\ref{eq52}) and Figure 7 lead to the $G_{\exp}$ value of $6.674\cdot 10^{-11}\;m^3kg^{-1}s^{-2}$ in quantitative agreement with experiment \cite{Mohr05,Gundlach00,Gillies97}, the selection of $m_{o}=(1/4)(\alpha /2\pi)^{3} m_{p}$ also leads to:

1. A $m_{o} $ value $(0.3675{\rm \; eV/c}^{{\rm 2}} )$ quite close to the estimated heaviest neutrino mass of $0.4{\rm \; eV/c}^{{\rm 2}} $ \cite{Griffiths08}.

2. A $V_{c} $ energy $(145.2{\rm \; MeV)}$ in good agreement with the QCD transition energy of $151(6)$ MeV \cite{Aoki06} and also with the energy 139.6 $MeV$, of $\pi^+$-mesons \cite{Griffiths08}.

\section{Consistency with quantum mechanics}

The above analysis was based entirely on special relativity and on the laws of Coulomb and Newton. Thus for completeness the results should be compared with quantum mechanics. 

As a first step it is interesting to note that the confinement distance $x_Tr_o$ of the two particles lies below or close to the proton Compton length, $\lambda _{c} $, and thus below or close to their de Broglie wavelength, which is something one normally would expect to find from quantum mechanics. Thus for the particle with $m_{o} =(1/4)(\alpha /2\pi )^{3} m_{o} $ it is $b=8/15$ and $\gamma _{p} =2(\alpha /2\pi )^{-3} $ (Fig. 7) and thus the particle momentum $p(x_{p} )$ at $x=x_{p} $ is given by: 
 
\begin{equation} \label{56)} 
p(x_{p} )=\gamma _{p} m_{o} c=2(\alpha /2\pi )^{-3} (1/4)(\alpha /2\pi )^{3} m_{p} c=(1/2)m_{p} c 
\end{equation}
 
Thus the particle de Broglie wavelength, $\lambda _{p} $, is given by 
 
\begin{equation} \label{57)} 
\lambda_{p}=\frac{h}{(1/2)m_{p} c} =2\lambda _{c}
\end{equation}

i.e., it coincides with the Compton wavelength of a particle with mass $m_p/2$, in consistency with the uncertainty principle.

The agreement between the maximum separation $x_{T} r_{o} $ and $\lambda _{c} $ becomes quantitative for $b=0.320816\approx 8/25$ (Fig. 10 which is based on Fig. 6b and eq. (\ref{eq41})). Very interestingly in this case it is $T_{osc} =8.81294\cdot 10^{-24} {\rm \; s}$ thus $f=1.1347\cdot 10^{23}\; s^{-1}$ and $\omega=2\pi f=7.1295\cdot 10^{23}s^{-1}$, therefore $hf=\hbar \omega =7.5186\cdot 10^{-11}J=469.274\;MeV=(1/2)m_{p} c^{2}$ (Fig. 6b and Fig. 10), which is the total energy, E, of each particle in the confined state. Thus in this case two equations follow:
 
\begin{equation} \label{eq58} 
h=(c/f)(m_pc/2)=\Delta r\Delta p
\end{equation}
 
where $\Delta r$ and $\Delta p$ express the uncertainty in distance and momentum due to the oscillations and:
 \begin{equation} \label{eq59} 
E=(1/2)m_pc^2=hf 
\end{equation}

The first equation (\ref{eq58}) shows the exact conformity of the solution, $f$, obtained from special relativity with Heisenberg's uncertainty principle and the de Broglie equation, while the second equation (\ref{eq59}) shows that, interestingly, the total energy of the oscillating particle is expressed in terms of $f$ by the Planck equation for the energy of a photon.
 
It is also interesting to note that the energy, $E$, corresponding to $b=8/15$, thus $T_{osc}=2.876\cdot10^{-26}\ s$ (Fig. 10) is $143.8\; GeV$ which lies within the Standard Model predictions about the mass $(115-180{\rm \; GeV/c}^{2})$ of the Higgs boson \cite{{Higgs64},{Yao06}}, but this may be a coincidence. 
\begin{figure}
\centerline{\includegraphics[width=11.75cm,height=8.50cm]{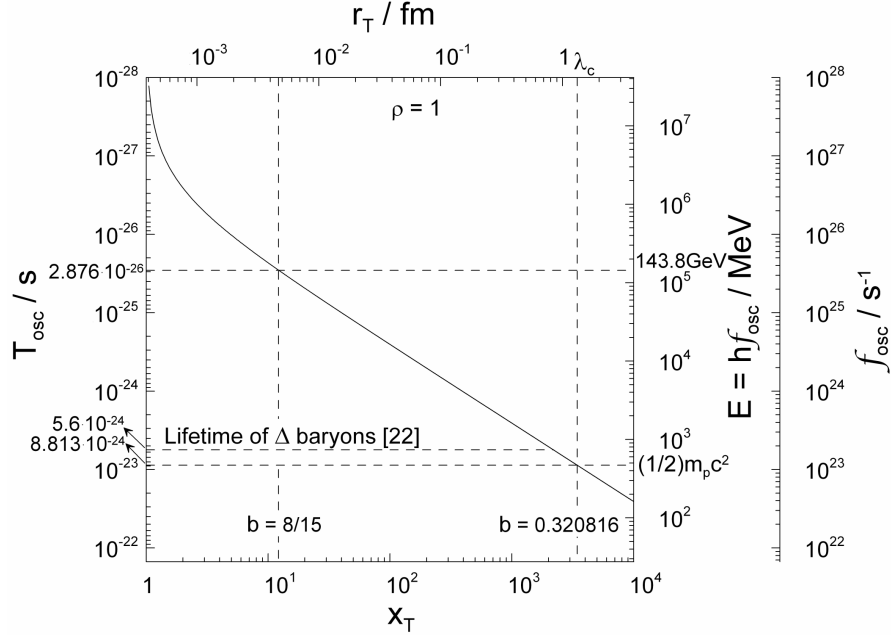}}
\caption{Effect of $x_{T}$ and oscillation amplitude $r_T$ on the period, frequency and energy of the oscillator.}
\label{fig10}
\end{figure}

Finally it is worth noting that in view of the very small value of $m_{o} c^{2} (\sim 0.4{\rm \; eV)}$ the $m_{o} c^{2} $ term in (\ref{eq8}) i.e. 
 
\begin{equation}
\label{eq60} 
E(x,t)=\sqrt{m_{o}^{2} c^{4} +p^{2} (x,t)\cdot c^{2} }  
\end{equation}
 
is negligible and thus it is $E(x,t)\approx p(x,t)\cdot c$ which implies that the time dependent Klein-Gordon equation reduces to the time-dependent Schr\"odinger equation:
 
\begin{equation}
\label{eq61} 
i\hbar \frac{\partial}{\partial t}\Psi (r,t)=\left(-\frac{\hbar ^{2} }{2m_{o}}\frac{\partial^{2}}{\partial x^{2} } +V(r)\right)\Psi (r,t) 
\end{equation}
 
which in turn reduces, due to the very small $m_{o} $ value, to:
 
\begin{equation} \label{eq62} 
i\hbar \frac{\partial }{\partial t} \Psi (r,t)=-\frac{\hbar ^{2} }{2m_{o} } \frac{\partial ^{2} \Psi (r,t)}{\partial x^{2}}
\end{equation}
 
which is the free-particle wave equation with solutions of the form \cite{Schwarz04}:
 
\begin{equation}
\label{eq63} 
\Psi (x,t)\sim e^{i(kr-\omega t)}
\end{equation}
 
where $k=\omega /c\approx 2.37\cdot 10^{15}$ m$^{-1}$, i.e. $k=2.37$ fm$^{-1}$ when using the above determined $\omega$ value. Thus in summary one may conclude that the creation of the bound oscillatory state by the two particles, obtained from special relativity, is also consistent with quantum mechanics.

\section{Gravitational confinement in circular trajectories}

The two light particles of mass $m_o$ forming bound hadron states need not be charged if their initial kinetic energy is sufficiently high. This point is worth analyzing, since neutrinos, e.g. electron neutrinos, have a charge radius $<r^2_{\nu_e}><3.32\cdot 10^{-36}m^2$ and $<|r_{\nu_e}|><1.83\cdot 10^{-3} fm$ [27,28]) but no net charge.

Thus if the two particles of rest mass $m_o$ have velocities v and -v relative to the laboratory observer and are moving initially in opposite directions on two parallel paths sufficiently close to each other, then , to a good approximation, the gravitational force judged by the laboratory observer is again:
\begin{equation}
\label{eq64}
F_G=\frac{Gm^2_o\gamma^6}{r^2}
\end{equation}
 
where r is again the distance and this force acting between the two particles becomes the centripetal force for their cyclic motion, i.e. [10]:

\begin{equation}
\label{eq65}
\frac{Gm^2_o\gamma^6}{r^2}=\frac{\gamma m_oc^2}{r/2}{\quad}; {\quad} r/2=\frac{Gm_o\gamma^5}{c^2}=\frac{Gm_n\gamma^4_p}{2c^2}
\end{equation}
 
where $m_n$ is the neutron mass and in the last equality we have set $\gamma=\gamma_p$ (which guaranties that $\mu=1$ and thus the confined mass is $m_n$) and have used $m_n=2\gamma_pm_o$ similarly to (\ref{eq7}). The problem described by (\ref{eq65}) is very similar to the Bohr treatment of the H atom with the gravitational attraction replacing the Coulombic attraction. 

In the linear motion problem we have chosen the parton mass from $m_o=(1/4n^2_m)(\alpha/2\pi)^3m_p$ with $n_m=1$. In the circular motion we choose $n_m=2$, thus $m_o=(1/16)(\alpha/2\pi)^3m_n$ and thus $\gamma_p=8(\alpha/2\pi)^{-3}$.

As in the Bohr treatment of the H atom we assume that the angular momentum is quantized, i.e. 
 
\begin{equation}
\label{eq66}
\gamma_pm_oc(r/2)=n\hbar
\end{equation}
 
and using (\ref{eq65}) to express $r/2$ and accounting for $\gamma_p=8(\alpha/2\pi)^{-3}$ one obtains:
 
\begin{equation}
\label{eq67}
G=\frac{4n\hbar c(\alpha/2\pi)^{12}}{8^4m_n^2} \quad thus \ for \  n=1 \quad G=6.6293\cdot10^{-11}\ m^{3}kg^{-1}s^{-2}
\end{equation}
 
which very interestingly, differs less than $0.1 \%$ from the value computed in (\ref{eq51}), which  involves $m_p$ instead of $m_n$, and less than $0.4\%$ from the experimental G value of $6.6742\cdot10^{-11}\ m^3kg^{-1}s^{-2}$ [19,20,21].

The mass $m_o$ of each one of the rotating particles is $m_o=(1/16)(\alpha/2\pi)^3m_p=0.0919$ $eV/c^2$ which is within the range of neutrino masses [22]. The distance $r$ of the two particles, i.e. the diameter of rotation, is obtained from (\ref{eq65}), i.e. 
 
\begin{equation}
\label{eq68}
r=\frac{Gm_n8^4(\alpha/2\pi)^{-12}}{c^2}=0.844 \ fm
\end{equation}
 
which is in excellent agreement with the estimated proton or neutron diameter $\lambda_c=1.32 \;fm$. The period of the rotation is $\pi r/c=8.838\cdot10^{-24}\; s$, i.e. almost identical with the period $8.813\cdot10^{-24}$ s computed for the linear motion with $r_ox_T=\lambda_c$ (Fig. 6b) and very close to the lifetime of $\Delta$ baryons $(5.6\cdot 10^{-24}$ s, Fig. 6b). Thus one may safely conclude that gravity can confine both charged and neutral particles of the mass range of neutrinos to form hadrons. The same analysis presented in this section also applies to pairs of charged and uncharged light particles (e.g. pairs with charges e and zero) and this eliminates the necessity of assuming fractional charges $Q_1, Q_2$ (e.g. 1/2 or $\pm 2/3$ and $\pm 1/3$) as in the linear motion case to form charged hadrons. 

\section{Discussion and conclusions}

 The analysis of the one-dimensional motion of two charged particles under the influence of their Coulombic repulsion and gravitational attraction using the relativistic equation of motion has shown that when the particles are sufficiently light $(m_{o} <0.4183{\rm \; eV/c}^{{\rm 2}}$, in the mass range of neutrinos) then gravity is sufficient to create a stable electrostatic-gravitational oscillator, i.e. a bound state, having the mass of a proton. This is also consistent with the de Broglie wavelength expression and with the Klein-Gordon and Schr\"odinger equations of quantum mechanics.

 The transformation of Coulombic energy into kinetic energy of the two particles and thus into rest energy R (and rest mass $m=R/c^{2} $) of the confined state, provides a straightforward mechanism of ``hadronization'' [22], i.e. of generation of rest mass when a bound state is formed by two fast particles.  

 Also when the initial kinetic energy of two neutral particles (neutrinos with mass 0.0919 $eV/c^2$) is sufficiently high $(m_pc^2)$, gravity alone suffices to confine these particles in stable circular (or elliptical) trajectories corresponding to neutral hadrons, e.g. neutrons.
 The relevance of the present simple analysis to the actual problem of hadron formation via the condensation of quark-gluon plasma [13, 16-18] appears worth further investigation for several reasons:

\begin{enumerate}
\item  It is interesting and reassuring that the computed maximum particle mass $m_{o,c} \simeq 0.4183{\rm \; eV/c}^{{\rm 2}} $ necessary for confinement lies in the range of the heaviest neutrinos $(\sim 0.4{\rm \; eV/c}^{{\rm 2}} )$. This shows that such small particles as the ones used in the present analysis, actually exist and in fact have velocities near the speed of light \cite{Griffiths08}. It is also worth reminding that during the decay of muons two, rather than one, neutrinos are produced \cite{Griffiths08}.

\item  The two fast moving (or rotating) particles appear to have the basic properties of gluons, i.e. they induce the strong force and they are practically massless \cite{Griffiths08, Nambu84, Povh06, Kronfeld08,Durr08}. Thus the formation of neutral hadrons such as neutrons, appears to follow logically from the present analysis. For the case of charged hadrons it follows that if one may view gluons as charged neutrinos, then the creation and stability of protons could also be readily rationalized on the basis of the present analysis. As already noted, in the case of circular trajectories the necessity of assuming integer or fractional charges is eliminated.

\item  The massive particles formed during each oscillation (as $\mu (x)$ oscillates and thus the rest mass of the system oscillates between $2m_{o}$ $(\sim 0.8 eV/c^2)$ and $2\gamma m_{o}$ $(>932 MeV/c^2)$) appear to have some of the basic properties of sea quarks \cite{Griffiths08}, i.e. they are produced and annihilated as virtual particles during each oscillation, exactly as envisioned in the standard model \cite{Griffiths08}. The maximum period of these oscillations $(\sim 8.7\cdot 10^{-24}\; s)$ lies very close to the lifetime $(\sim 5.6\cdot 10^{-24}\; s)$ of $\Delta $ baryons \cite{Griffiths08}. 

\begin{figure}
\centerline{\includegraphics[width=8.75cm,height=7.80cm]{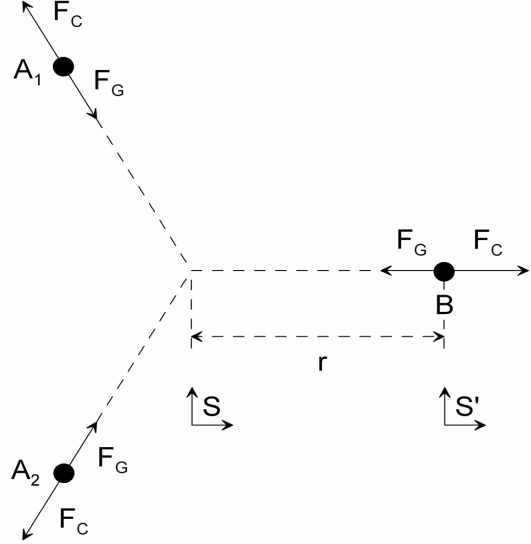}}
\caption{Geometry of the one dimensional motion of particle B involving two other Coulombically repelling charged particles A1 and A2 replacing particle A in Figure 1 in a symmetric arrangement. The two reference frames $S$ and $S'$ remain the same.}
\label{fig11}
\end{figure}

In this direction it is worth pointing out that the present analysis of the motion of particle B is directly applicable, with very minor modifications, to the motion of the same particle in presence of two other particles in a symmetric arrangement (Fig. 11), and this geometry, to which the present results are directly applicable, involving three particles, is very close to the standard model picture of three quarks produced and annihilated continuously inside a proton \cite{Griffiths08}.

\item  The Coulombic and the negative of the gravitational potential energy at the confinement point $x_{p} $ are equal to $145.7{\rm \; MeV}$, very close to the energy $151(6)$ MeV of quarks and gluons at the condensation point of quark-gluon plasma to form hadrons \cite{Braun07, Cabibbo75, Aoki06, Fodor04} and also close to the energy, 139.6 $MeV$, of $\pi^+$-mesons \cite{Griffiths08}.

\item  The limiting (for $r\to \infty $) escape rest mass $R/c^2$ for $b<8/25$ (e.g. $b=2/9$, Fig. 3a) is very near the rest masses of all the $(1/2)$ spin hadrons, i.e. the $\xi ,{\rm \; }\Sigma ^{{\rm +}} ,{\rm \; }\Sigma ^{{\rm o}} $ and $\Sigma ^{-} $ baryons \cite{Griffiths08} (Fig. 3a).

\item  The rest mass $R/c^{2}$ for $b>8/25$ (e.g. $b=8/15$, Fig. 3b) during the oscillations between $x_{p} $ and $x_{T} $ covers the range of the rest masses of all the $(3/2)$ spin baryons, i.e. the $\Delta ,{\rm \; }\Sigma ^{{\rm *}} ,{\rm \; }\Xi ^{*} $ and $\Omega^{^{-}}$ baryons (Fig. 3b).

\item  The results are consistent with quantum mechanics, and when $r_ox_T=\lambda_c$, then the energy $\hbar \omega$ equals $(1/2)m_{p}c^{2}$ (Figs. 6b and 10). 

\item Finally the analysis provides a straightforward explanation about why individual quarks or gluons cannot be separated from hadrons or mesons and studied independently. Although the analysis shows that, at first surprisingly, the strong force can be viewed as the relativistic gravitational force, the emerging picture is very similar to that of the standard model regarding the strong force and the composition of hadrons \cite{Griffiths08,Nambu84, Kronfeld08, Durr08} as well as their behaviour in elastic and inelastic scattering experiments which has led to the concepts of partons connected by strings \cite{Nambu84}, of the bag model \cite{Povh06} and of the massive quarks and practically massless gluons \cite{Griffiths08,Nambu84, Kronfeld08, Durr08}. All these concepts are consistent with the present analysis at different phases of the oscillation.
\end{enumerate}

Regardless of the exact direct or indirect relevance of the present simple analysis to the above important physical phenomena, particles, and concepts it is certain that gravitational forces suffice to create stable electrostatic-gravitational oscillator states and this leads in a straightforward manner to simple formulae (\ref{eq51} to \ref{eq53} and \ref{eq67}) for the gravitational constant and for the proton mass in terms of the other physical constants which are in quantitative agreement with experiment.

It is also reasonable to expect that the gravitational forces exerted between the fast moving particle constituents (partons or quarks) of neighboring hadrons at ${\rm fm}$ distances can also be quite important and thus can lead to binding energies per nucleon of the order of $\alpha m_{p}c^{2} \approx 7{\rm \; MeV}$ \cite{Griffiths08} and the formation of nuclei. Preliminary work \cite{Vayenas07, Vayenas2007} has shown that indeed the binding energies of some light nuclei ($^{2}H$ and $^{4}He$) can thus be computed using (\ref{eq51}) with good accuracy and this point also appears to deserve further investigation.

\section*{Acknowledgment}
\noindent CGV acknowledges helpful discussion with Professors Ilan Riess and Amos Ori from the Physics Department of the Technion and with Dr. E. Riess in the summer of 2008.

\newpage
\bibliography{bib}

\newpage
\begin{tabular}{p{12.0cm}p{0.0cm}}
\multicolumn{2}{p{12cm}}{\textbf{Table 1. CODATA recommended values of $e$, $\varepsilon$, $m_{p} $ and G and comparison of the value of $\xi (=\varepsilon Gm_{p}^{2} /e^{2} )$, $(=\xi _{\exp})$ computed from them and from equations (\ref{eq50}), (\ref{eq51}) and (\ref{eq52})}} \\
$e=1.6021765\times 10^{-19} {\rm \; C}$ \\ 
$\varepsilon =4\pi \varepsilon _{o} =1.112649 \times 10^{-10} {\rm \; C}^{2}/Nm^{2}$\\ 
$h=6.6260693\times10^{-34}\ Js$\\
$c=2.997925 \times 10^{8} \ ms^{-1}$\\
$m_{n}=939.565\ MeV/c^{2}=1.67492728 \times 10^{-27} \ kg$\\
$m_{p}=938.2723\ MeV/c^{2}=1.67262171 \times 10^{-27} \ kg$\\
$G_{\exp}=6.6742\times 10^{-11} {\rm \; m}^{{\rm 3}} kg^{-1} s^{-2}$\\ 
$\xi _{\exp } =8.093421\times 10^{-37} =0.1343595\times (\alpha /2\pi )^{12}$\\
$\xi =8.031608\times 10^{-37} =0.13333\times (\alpha /2\pi)^{12}$\\
$G=6.62318\times 10^{-11} {\rm \; m}^{{\rm 3}} kg^{-1}s^{-2}=0.13333(\alpha/2\pi )^{12} (e^{2}/\varepsilon m_{p}^{2})$\\
$G=6.674\times 10^{-11} m^{3} kg^{-1} s^{-2} =0.13333(\alpha /2\pi )^{12} (e^{2}/\varepsilon m_{p,m}^{2})$\\
$m_{p,m}$ in $^{56}Fe=934.547{\rm \; MeV/c}^{2}$ and in $^{98}Mo=934.647{\rm \; MeV/c}^{2}$\\
$m_{p,m}$ from eq. (\ref{eq53}): $m_{p,m} =934.6{\rm \; MeV/c}^{2}$\\
$\alpha=1/137.035$\\ 
$(\alpha /2\pi)=e^{2}/\varepsilon ch=1.16141822\times 10^{-3}$\\ 
$(\alpha /2\pi)^{3}=1.5666281\times 10^{-9}$\\
$(\alpha /2\pi )^{12} =6.02370395\times 10^{-36} $\\
  $(\alpha/2\pi)^{-12}=1.6601081\times 10^{35}$\\
\end{tabular}
 
\newpage
\textbf{Table 2. List of symbols}
\newline
\begin{tabular}{p{4.0in}}
$b:{\rm \; \; }(\varepsilon Gm_{p}^{2}/16Qe^{2}) \gamma_{p}^{4}=\xi(\gamma_{p}^{4}/16Q)$\\
$m_{\ell } :{\rm \; \; particle\; longitudinal\; mass,\; }\gamma ^{{\rm 3}}m_{o}$\\
$m_{o} :{\rm \; \; particle\; rest\; mass}$\\
$m_{n} :{\rm \; \; neutron\; mass}$\\
$m_{p} :{\rm \; \; proton\; mass}$\\
$m_{p,m}$: average nucleon mass in a nucleus \\
$n,n_{m}$: positive integers \\
$p:{\rm \; \; momentum}$\\
$q_{1},\; q_{2}\;:{\rm \; particle\; charges}$\\
$Q_{1} ,{\rm \; Q}_{{\rm 2}} :{\rm \; }q_{1} /e,{\rm \; }q_{2} /e$\\
$Q:{\rm \; \; }Q_{1} {\rm Q}_{{\rm 2}} $\\
$x:{\rm \; dimensionless\; distance\; r}/{\rm r}_{{\rm o}} $\\
$x_{m} :{\rm \; value\; of\; x\; at\; maximum\; particle\; momentum}$\\
$x_{p} :{\rm \; value\; of\; x\; at\; }\mu {\rm (x)}={\rm 1}$\\
$x_{T} :{\rm \; terminal\; x\; value\; during\rm \; oscillations}$\\
$y(x):{\rm \; \; }-V_{G} (x)/V_{C,o} $\\
\textbf{Greek symbols} \\
$\alpha :{\rm \; \; e}^{{\rm 2}} /\varepsilon c\hbar $\\
$(\alpha /2\pi ):{\rm \; \; e}^{{\rm 2}} /\varepsilon ch$\\
$\beta :{\rm \; \; v/c}$\\
$\gamma :{\rm \; \; }(1-v^{2} /c^{2} )^{-1/2} $\\
$\gamma _{p} :{\rm \; value\; of\; }\gamma {\rm (x)\; at\; x}={\rm x}_{{\rm p}} $\\
$\lambda _{c} =h/m_{p}c$\\
$\lambda _{q} =e^{2} /\varepsilon (m_{p} /2)c$\\
$\mu (x):{\rm \; \; dimensionless\; mass}=\gamma (x)/\gamma _{p} =R(x)/m_{p} $\\
$\xi :{\rm \; \; }(\varepsilon {\rm Gm}_{{\rm p}}^{{\rm 2}}/{\rm e}^{{\rm 2}}) $\\
$\rho :{\rm \; \; V}_{{\rm C,o}} /m_{p} c^{2} =Qe^{2} /\varepsilon r_{o} m_{p} c^{2}$\\
\end{tabular}
 
\newpage
\textbf{Table 3.} Summary of key mathematical equations describing the energy balance and the particle equation of motion

\renewcommand{\arraystretch}{3.0}
\begin{tabular}{|c|c|c|}
\cline{2-3}
\multicolumn{1}{c|}{} & Case A $(2K_{o}=(1-\rho)m_{p}c^{2})$ & Case B $(K_{o} =0)$\\ \hline \cline{2-3}
\multicolumn{1}{|c|}{\multirow{3}{3.5cm}{\small{Energy balance and corresponding gravitational potential energy profile\\ $y(x)=-\frac{(V_{G}(x)-V_{G,o})}{V_{C,0}}$}}} & \multicolumn{1}{|c|}{$\frac{dy(x)}{dx}=4b\frac{(1-\rho /x + \rho y(x))^{6} }{x^{2}}$\;(\ref{eq22})} & $\frac{dy(x)}{dx} =4b\rho ^{6} \frac{(1-1/x+y(x))^{6} }{x^{2}}\; (\ref{eq23})$ \\ \cline{2-3}
 & \multicolumn{2}{|c|}{$BC:\ y(1)=0$} \\ \cline{2-3}
 & \multicolumn{1}{|c|}{$\mu (x)=1-\rho \left((1/x)-y(x)\right)$\;(\ref{eq24})} & $\mu (x)=\rho \left(1-1/x+y(x)\right)$\;(\ref{eq25}) \\ \hline
\multicolumn{1}{|c|}{\multirow{2}{*}{\small{Equation of Motion}}} & \multicolumn{2}{|c|}{$\frac{dp}{dt} =F_{C} (x)\left[1-4b\mu ^{6} (x)\right]=(2\rho^{2}m_{p}c^{2}/Q\lambda_{q}x^{2})\left[1-4b\mu ^{6} (x)\right]$\;(\ref{eq34}),\;(\ref{eq28})} \\  \cline{2-3}
 & \multicolumn{2}{|c|}{$T_{osc} \approx (\lambda_{q}/{4\rho c})(x_{T}-1)$\;(\ref{eq40}),\;(\ref{eq41})} \\ \hline
\end{tabular}
\end{document}